\documentclass[%
 reprint,
 amsmath,amssymb,
 aps,
]{revtex4-1}
\usepackage{float}
\usepackage{graphicx}% Include figure files
\usepackage{dcolumn}% Align table columns on decimal point
\usepackage{bm}
\usepackage[utf8]{inputenc}
\usepackage{graphicx}
\usepackage{colortbl,array} % für farbige cells
\usepackage{multirow,bigdelim}
\usepackage{booktabs}
\usepackage[flushleft]{threeparttable}
\usepackage{float}
\usepackage{caption}
\usepackage{subcaption}
\usepackage{bigdelim}
\usepackage{tabularx}
\usepackage{amsmath}
\usepackage{multirow}
\usepackage{amssymb,bm,mathrsfs,bbm,amscd}
\usepackage{ragged2e}
\usepackage{hhline}
\usepackage{tabularx}
\usepackage{caption}
\usepackage{multirow}
\usepackage{placeins}
\usepackage{array}
\usepackage{booktabs}
\makeatletter
\newcommand{\thickhline}{%
    \noalign {\ifnum 0=`}\fi \hrule height 1pt
    \futurelet \reserved@a \@xhline
}
\newcolumntype{"}{@{\hskip\tabcolsep\vrule width 1pt\hskip\tabcolsep}}
\makeatother
\def\be{\begin{equation}}
\def\ee{\end{equation}}
\begin{document}

\preprint{APS/123-QED}

\title{Description of moment of inertia and the interplay between anti-pairing and pairing correlations in
even-even $^{244}$Pu and $^{248}$Cm }% Force line breaks with \\

\author{Anshul Dadwal}
\email{dadwal.anshul@gmail.com}
 %\altaffiliation[Also at ]{Dr. B. R. Ambedkar National Institute of Technology,
     %Jalandhar 144011, India.}%Lines break automatically or can be forced with \\
\author{Xiao-Tao He}%
 %\email{mittal.hm@lycos.com}
\affiliation{%
Department of Nuclear Science and Technology,\\
College of Materials Science and Technology,\\
Nanjing University of Aeronautics and Astronautics, Nanjing 210016, China.
}%


%\affiliation{
%\affiliation{
%\affiliation{%

\date{\today}% It is always \today, today,

\begin{abstract}
Within the supersymmetry scheme, which includes many-body interactions and a perturbation possessing the SO(5) (or SU(5)) symmetry, the rotational bands of the $A\sim 250$ mass region are studied systematically. A novel modification is introduced,
extending the Arima coefficient to the third order. This study is dedicated to the quantitative  analysis of evolving trends in intraband $\gamma$-transition energy, kinematic, and the dynamic moment of inertia within the rotational bands of $^{244}$Pu and $^{248}$Cm. The computed outcomes exhibit an exceptional degree of agreement with experimental observations across various conditions. The significance of including a higher-order Arima coefficient is further examined by contrasting it with the previously suggested model. The calculated results demonstrate the significance of both the anti-pairing and pairing effects in the evolution of the dynamic moment of inertia. Additionally, these insights reveal the importance of a newly introduced parameter in accurately depicting complex nuclear behaviors such as back-bending, up-bending, and the downturn in the moment of inertia.
\begin{description}
\item[PACS numbers]
21.10.Hw, 21.10.Re, 21.60.Ev.
\end{description}
\end{abstract}

\pacs{Valid PACS appear here}% PACS, the Physics and Astronomy
\maketitle

%\tableofcontents
% ===================================================================
\section{Introduction}

One of the intriguing puzzles within the field of nuclear structure physics pertains to the single-particle composition of heavy and superheavy elements (SHE). This inquiry poses a significant challenge for both theoretical frameworks and experimental investigations. Experimental endeavors predominantly concentrate on determining the specific location and scope of the ``island of stability". The emergence of SHE is attributed to shell effects, as the liquid-drop model predicts the non-existence of such nuclei due to substantial Coulomb repulsions. An essential question in this pursuit involves the exploration of magic numbers beyond $Z=82$ and $N=126$ in SHE, representing a crucial aspect for both theoretical frameworks and experimental inquiries. The identification of new magic numbers is intricately linked to the single-particle structure. Theoretical predictions propose that nuclei in close proximity to N=184 and Z=114 may indicate the presence of an island of stability \cite{Stoyer2006}. Currently, the availability of detailed spectroscopy data for $Z\approx100$ opens up a new realm for systematically studying the evolution of single-particle states \cite{nndc}.
The exploration of super-heavy elements (SHE) is constrained by the narrow cross-sections, resulting in a scarcity of experimental data to corroborate theoretical predictions. The experimental approaches for studying SHE can be broadly categorized into two types: in-beam and decay spectroscopy \cite{HerzbergCox+2011+441+457}. The in-beam spectroscopic technique is pivotal in examining rotational bands and the single-particle structure, even for faint channels. Despite its complexity, in-beam conversion electron spectroscopy is profoundly valuable, as it establishes rotational bands and extracts crucial information about the alignments of protons and neutrons, even with just a few dozen gamma rays.
Decay spectroscopy is instrumental in analyzing single-particle levels through alpha decay chains. Alpha decay in odd-mass nuclei typically removes pairs of protons and neutrons, leaving behind an unpaired nucleon in the mother nucleus. The state populated in the daughter nucleus and the ground state of the mother nucleus possess the same single-particle configuration. The daughter nucleus's excited state decays to the ground state, emitting secondary gamma rays and conversion electrons, which aids in determining the excitation energy of single-particle states.
At present, the most substantial spectroscopic information available is for the heaviest nuclei, particularly the transfermium elements such as californium, fermium, and nobelium \cite{Leino,Herzberg_2004,GREENLEES2007507,HERZBERG2008674}. Although these deformed nuclei, with $Z\approx100$ and $N\approx150-160$, are not strictly classified as SHE, they represent a threshold to the SHE region. The increasing sensitivity of experimental setups at facilities like ANL (Argonne), GSI (Darmstadt), JYFL (Jyväskylä), GANIL (Caen), and FLNR (Dubna) has made it feasible to measure $\alpha-\gamma$ or $\alpha$ conversion-electron coincidences \cite{Reiter1,Reiter2,Reiter3,Hump,Butler,Herzberg,Bastin}.\par
The one of the most essential quantity is moment of inertia (MoI) which characterizes the nuclear rotational bands. The MoI
is vastly studied and it is the most fundamental observable to illustrate the structure of the nuclei.
To describe the high spin phenomena of the rotational bands, primarily, two types of the MoI are given, i.e. the
kinematic ($\Im^{(1)}$) and dynamic ($ \Im^{(2)}$) MoI.
The calculation of dynamic moment of inertia (MoI) is advantageous over the kinematic MoI, as it does not necessitate knowledge of the
spins. Systematic studies of MoI have uncovered some remarkable features in the $A\sim250$ mass region. Specifically, an investigation
into the MoI systematics of plutonium isotopes reveals distinct behaviors between lighter isotopes ($A\sim 238-240$) and heavier ones
($A\geq241$). The lighter isotopes do not exhibit the upbending in MoI that is observed in the heavier counterparts
\cite{PhysRevLett.83.2143}. In the lightest isotopes, strong octupole correlations are present, which are believed to be responsible
for the absence of significant proton alignment, a feature observed in heavier Pu isotopes \cite{PhysRevLett.83.2143}. For $^{240}$Pu,
the lack of alignment has been interpreted in terms of phonon condensation \cite{PhysRevLett.102.122501}. In the case of Cm isotopes,
the dynamic MoI displays an extraordinary pattern: initially, there is a smooth up-bend followed by a downturn in the dynamic MoI in
the ground band. This behavior was suggested to be a result of the interplay between $j_{15/2}$ neutrons and $i_{13/2}$ protons
\cite{PhysRevLett.46.415,RBPiercey1993}.\par
%%%%%%%%%%%%%%%%%%%%%%%%%%%%%%%%%%%%%%%%%%%%%%%%%%%%%%%%%%%%%%%%%%%%%%%%%%%%%%%%%
In the present paper, we have systematically studied the rotational bands in the $^{244}$Pu and $^{248}$Cm with the perturbed
$SU_{sdg}(3)$ symmetry with the perturbation holding the $SO_{sdg}(5)$ symmetry which is very successful in reproducing the changing
behaviour of dynamic MoI \cite{Yuxin_JPG,yuxin_prc_1,yuxin_prc_2,yuxin_prc_3,yuxin_prc_4,yuxin_prc_5,yuxin_prc_6}. In section II, a
short description of the model is presented and an extension to the previous model is proposed via incorporating another coefficient in the previously defined Arima coefficient. In section III, the calculated results are presented for even-even $^{244}$Pu and $^{248}$Cm nuclei. Finally, summary is given in section IV.
% ===================================================================
\section{Formalism}
% ===================================================================

The Hamiltonian of the variable moment of inertia (VMI) inspired interacting boson model (IBM) is \cite{Yuxin_JPG}
\begin{equation}
H=E_{0}+\kappa \hat{Q}^{(2)}.\hat{Q}^{(2)}+\frac{C_{0}}{1+f\hat{L}.\hat{L}}\hat{L}.\hat{L}
\label{eq1}
\end{equation}
where $\hat{Q}^{(2)}$ and $ \hat{L}$ are the quadrupole and angular momentum operator, respectively. The parameter $f$ is known as the Arima coefficient which was introduced in IBM-1. The parameter $f$ is a spin dependent term which was introduced in the denominator of Hamiltonian to increase the moment of inertia. It was proposed that the spin dependent term $f\hat{L}.\hat{L}$ term in IBM Hamiltonian include the anti-pairing effect at high spins in the phenomenological studies \cite{YOSHIDA}. However, in the superdeformed bands (SD) of the $A\sim150,190$ mass region, a turnover is the dynamic MoI is observed. It was emphasised that the extending Arima coefficient is important to describe the changing feature of the dynamic MoI. Following the VMI model, Arima coefficient $f$ was extended as $f=f_1+f_2 [I(I+1)]$. Hence the energy expression in the framework of the VMI model can be written as \cite{Yuxin_JPG}
\begin{align}\label{eq2}
  E= & E_{0}(N_{B},N_{F}) \nonumber\\[1mm]
   & +\frac{C_{0}}{1+f_{1}I(I+1)+f_{2}I^{2}(I+1)^{2}}I(I+1).
\end{align}
Here, $N_B$ and $N_F$ represents the boson and fermion numbers, respectively. This expression is the one given by a core
with the $SU(3)$ symmetry plus a pseudospin $S$. The \ref{eq2} generates the rotational band which reproduced the global turnover of dynamic MoI pretty well. However, the calculated dynamic MoI changes very smoothly with rotational frequency that the weak $\Delta I=2$ staggering is completely ignored. To describe the $\Delta I=4$ bifurcation, the SU(3) symmetry must be broken and the interaction $SU_{sdg}(5)$ symmetry as a perturbation was taken into account \cite{yuxin_prc_1}. Hence, the energy of the state can be written as
\begin{align}
    E &= E_{0}(N_{B}, N_{F}) \nonumber \\[2mm]
    &\quad + A \biggl[ n_{1}(n_{1}+4) + n_{2}(n_{1}+2) \nonumber \\[2mm]
    &\quad + n_{3}^{2} + n_{4}(n_{4}-2) - \frac{1}{5}(n_{1}+n_{2}+n_{3}+n_{4})^{2} \biggr] \nonumber \\[2mm]
    &\quad + B[\tau_{1}(\tau_{1}+3) + \tau_{2}(\tau_{2}+1)] \nonumber \\[2mm]
    &\quad + \frac{C_{0}}{1+f_{1}I(I+1)+f_{2}I^{2}(I+1)^{2}} I(I+1).
    \label{eq3}
\end{align}
The perturbed SU(3) limit of the $sdg$ IBM can describe the rotational bands. Moreover, the $SU_{sdg}(5)$ limit of $sdg$ IBM is relevant for deformed nuclei as well as the $SU_sdg$(3) limit. The calculation of hexadecupole deformation parameter $\beta_{4}$, the two-nucleon transfer cross section, and the energy spectra illustrated that $SU_{sdg}(5)$ limit has almost the same property in
describing deformed rotational nuclear spectra as the $SU_{sdg}(3)$ does. This implies that the $SU_{sdg}(5)$ symmetry of the $sdg$ IBM incorporates a shape coexistence and shape phase transformation which is directed by the hexadecupole deformation and angular momentum.
Since the irreducible representation (irrep) ($\lambda,\mu$), the irrep $[n_{1},n_{2},n_{3},n_{4}]$ of $SU_{sdg}(5)$ contributes nothing to the excitation energy of the states in the band. Hence, only the contribution of the perturbation to the energy of the SD bands is with the $SO_{sdg}(5)$ symmetry. Now the Eq.(\ref{eq3}) can be written as
\begin{align}
E & =E_{0}(N_{B},N_{F})\nonumber\\[1mm]
& \qquad +B[\tau_{1}(\tau_{1}+3)+\tau_{2}(\tau_{2}+1)]\nonumber\\[1mm]
 & \qquad +\frac{C_{0}}{1+f_{1}I(I+1)+f_{2}I^{2}(I+1)^{2}}I(I+1)
\label{eq4}
\end{align}
where $I=I-i,(\tau_{1},\tau_{2})$ is the irrep of the SO(5) group.
In more realistic calculation, the irrep $(\tau_{1},\tau_{2})$ are given as
\begin{equation}
(\tau_{1}, \tau_{2}) =
\begin{cases}
    \begin{aligned}
        &\biggl(\dfrac{L}{2}, 0\biggr), \\
        &\text{if } L = 4k, 4k + 1 \quad (k = 0, 1, \ldots)
    \end{aligned} \\
    \begin{aligned}
        &\biggl(\dfrac{L}{2} - 1, 2\biggr), \\
        &\text{if } L = 4k + 2, 4k + 3 \quad (k = 0, 1, \ldots)
    \end{aligned}
\end{cases}
\label{eq5}
\end{equation}
Here $[L/2]$ denoted the integer part of the $L$ and $B$, $C_{0}$, $f_{1}$ and $f_{2}$ are the free parameters.
Building upon the foundational concepts presented in previous studies \cite{YOSHIDA,Yuxin_JPG,yuxin_prc_1,yuxin_prc_2,yuxin_prc_3,yuxin_prc_4,yuxin_prc_5,yuxin_prc_6,ZhangDaLi,ZhangDaLi2}, which advocate for the necessity of higher-order terms in Arima coefficients to incorporate effects that either promote pairing favouring and anti-pairing favouring effect, this research introduces an additional independent variable, $f_3$, to more accurately represent the rotational bands within the nuclear mass region around $A\sim250$. Consequently, the formulation of the Arima coefficient $f$ is revised to encompass a more complex structure, expressed as $f=f_1+f_2 [I(I+1)]+f_3 [I(I+1)]^2$. Utilizing this refined approach, key nuclear properties such as the energy of $E_2$ transition $\gamma$-rays, kinematic, and dynamic MoI are calculated with enhanced precision.

\begin{table*}
\caption{The parameters obtained using least-squares fitting method for rotational bands in $^{244}$Pu and $^{248}$Cm.
B and $C_0$ are in keV and $\chi$ represents the RMS-deviation between calculated and experimental $E_\gamma$ transitions.
Here $1,2,..$ in the parenthesis represent band 1, band 2...,respectively.}\label{tb1}
\begin{tabular}{c@{\hspace{1.5em}}c@{\hspace{1.5em}}c@{\hspace{1.5em}}c@{\hspace{1.5em}}c@{\hspace{1.5em}}
c@{\hspace{1.5em}}c@{\hspace{1.5em}}c@{\hspace{1.5em}}}
\thickhline\\[0.1cm]
Nucleus(Band) &Set	& B &	$C_{0}$	&	$f_1$&	$f_2$ &	$f_3$&	$\chi \times 10^{-3}$\\[0.1cm]
\hline\noalign{\smallskip}\\[0.1cm]
%% ================================================================================================
 \kern-\nulldelimiterspace
  \begin{tabular}{@{}c@{}c@{}}
                  \\[0.1cm]
    $^{244}$Pu(1) \\[0.1cm]
                  \\[0.1cm]
  \end{tabular}
 & $\kern-\nulldelimiterspace\left\{
  \begin{tabular}{@{}c@{}c@{}}
    A\\[0.1cm]
    B\\[0.1cm]
    C\\[0.1cm]
  \end{tabular}\right.$
  & $\kern-\nulldelimiterspace
  \begin{tabular}{@{}c@{}c@{}}
    0.0213\\[0.1cm]
    -0.0452	\\[0.1cm]
        0.0238\\[0.1cm]
  \end{tabular}$ &$\kern-\nulldelimiterspace
  \begin{tabular}{@{}c@{}c@{}}
    7.290\\[0.1cm]
    8.495\\[0.1cm]
    6.920    \\[0.1cm]
  \end{tabular}$ &$\kern-\nulldelimiterspace
  \begin{tabular}{@{}c@{}c@{}}
    $3.732\times10^{-4}$\\[0.1cm]
    $6.234 \times10^{-4}$\\[0.1cm]
        $2.682 \times10^{-5}$\\[0.1cm]
  \end{tabular}$ &$\kern-\nulldelimiterspace
  \begin{tabular}{@{}c@{}c@{}}
                         \\[0.1cm]
    $-1.073 \times10^{-7}$\\[0.1cm]
        $5.519 \times10^{-7}$ \\[0.1cm]
  \end{tabular}$ &$\kern-\nulldelimiterspace
  \begin{tabular}{@{}c@{}c@{}}
    \\[0.1cm]
    \\[0.1cm]
  $-2.608 \times10^{-10}$\\[0.1cm]
  \end{tabular}$ &$\kern-\nulldelimiterspace
  \begin{tabular}{@{}c@{}c@{}}
    46.6\\[0.1cm]
    49.5\\[0.1cm]
    29.8\\[0.1cm]
  \end{tabular}$ \\
%% ================================================================================================
%% ================================================================================================
 \kern-\nulldelimiterspace
  \begin{tabular}{@{}c@{}c@{}}
                  \\[0.1cm]
    $^{244}$Pu(2) \\[0.1cm]
                  \\[0.1cm]
  \end{tabular}
 & $\kern-\nulldelimiterspace\left\{
  \begin{tabular}{@{}c@{}c@{}}
    A:\\[0.1cm]
    B:\\[0.1cm]
    C:\\[0.1cm]
  \end{tabular}\right.$
  & $\kern-\nulldelimiterspace
  \begin{tabular}{@{}c@{}c@{}}
    -0.1865\\[0.1cm]
    -0.2680	\\[0.1cm]
    0.0784\\[0.1cm]
  \end{tabular}$ &$\kern-\nulldelimiterspace
  \begin{tabular}{@{}c@{}c@{}}
    6.845\\[0.1cm]
    7.362\\[0.1cm]
    4.923\\[0.1cm]
  \end{tabular}$ &$\kern-\nulldelimiterspace
  \begin{tabular}{@{}c@{}c@{}}
    $3.015\times10^{-4}$\\[0.1cm]
    $3.869 \times10^{-4}$\\[0.1cm]
    $-3.804 \times10^{-4}$\\[0.1cm]
  \end{tabular}$ &$\kern-\nulldelimiterspace
  \begin{tabular}{@{}c@{}c@{}}
                         \\[0.1cm]
    $-4.473 \times10^{-8}$\\[0.1cm]
        $9.289 \times10^{-7}$ \\[0.1cm]
  \end{tabular}$ &$\kern-\nulldelimiterspace
  \begin{tabular}{@{}c@{}c@{}}
    \\[0.1cm]
    \\[0.1cm]
  $-5.008 \times10^{-10}$\\[0.1cm]
  \end{tabular}$ &$\kern-\nulldelimiterspace
  \begin{tabular}{@{}c@{}c@{}}
    6.40\\[0.1cm]
    7.06\\[0.1cm]
    1.85\\[0.1cm]
  \end{tabular}$ \\
%% ================================================================================================
%% ================================================================================================
 \kern-\nulldelimiterspace
  \begin{tabular}{@{}c@{}c@{}}
                  \\[0.1cm]
    $^{244}$Pu(3) \\[0.1cm]
                  \\[0.1cm]
  \end{tabular}
 & $\kern-\nulldelimiterspace\left\{
  \begin{tabular}{@{}c@{}c@{}}
    A:\\[0.1cm]
    B:\\[0.1cm]
    C:\\[0.1cm]
  \end{tabular}\right.$
  & $\kern-\nulldelimiterspace
  \begin{tabular}{@{}c@{}c@{}}
    -0.1490\\[0.1cm]
    0.2047	\\[0.1cm]
    -0.1358\\[0.1cm]
  \end{tabular}$ &$\kern-\nulldelimiterspace
  \begin{tabular}{@{}c@{}c@{}}
    7.019\\[0.1cm]
    4.970\\[0.1cm]
    5.952\\[0.1cm]
  \end{tabular}$ &$\kern-\nulldelimiterspace
  \begin{tabular}{@{}c@{}c@{}}
    $2.343\times10^{-4}$\\[0.1cm]
    $-2.098 \times10^{-4}$\\[0.1cm]
    $5.952 \times10^{-4}$\\[0.1cm]
  \end{tabular}$ &$\kern-\nulldelimiterspace
  \begin{tabular}{@{}c@{}c@{}}
                         \\[0.1cm]
    $2.899 \times10^{-7}$\\[0.1cm]
        $-8.602 \times10^{-7}$ \\[0.1cm]
  \end{tabular}$ &$\kern-\nulldelimiterspace
  \begin{tabular}{@{}c@{}c@{}}
    \\[0.1cm]
    \\[0.1cm]
  $6.767 \times10^{-10}$\\[0.1cm]
  \end{tabular}$ &$\kern-\nulldelimiterspace
  \begin{tabular}{@{}c@{}c@{}}
    7.07\\[0.1cm]
    2.49\\[0.1cm]
    0.91\\[0.1cm]
  \end{tabular}$ \\
%% ================================================================================================
%% ================================================================================================
 \kern-\nulldelimiterspace
  \begin{tabular}{@{}c@{}c@{}}
                  \\[0.1cm]
    $^{244}$Pu(4) \\[0.1cm]
                  \\[0.1cm]
  \end{tabular}
 & $\kern-\nulldelimiterspace\left\{
  \begin{tabular}{@{}c@{}c@{}}
    A:\\[0.1cm]
    B:\\[0.1cm]
    C:\\[0.1cm]
  \end{tabular}\right.$
  & $\kern-\nulldelimiterspace
  \begin{tabular}{@{}c@{}c@{}}
    -0.1047\\[0.1cm]
    0.001	\\[0.1cm]
    -0.1912\\[0.1cm]
  \end{tabular}$ &$\kern-\nulldelimiterspace
  \begin{tabular}{@{}c@{}c@{}}
    6.641\\[0.1cm]
    5.968\\[0.1cm]
    7.310\\[0.1cm]
  \end{tabular}$ &$\kern-\nulldelimiterspace
  \begin{tabular}{@{}c@{}c@{}}
    $1.916\times10^{-4}$\\[0.1cm]
    $-5.790 \times10^{-5}$\\[0.1cm]
    $5.267 \times10^{-4}$\\[0.1cm]
  \end{tabular}$ &$\kern-\nulldelimiterspace
  \begin{tabular}{@{}c@{}c@{}}
                         \\[0.1cm]
    $2.227 \times10^{-7}$\\[0.1cm]
        $-6.626 \times10^{-7}$ \\[0.1cm]
  \end{tabular}$ &$\kern-\nulldelimiterspace
  \begin{tabular}{@{}c@{}c@{}}
    \\[0.1cm]
    \\[0.1cm]
  $4.978 \times10^{-10}$\\[0.1cm]
  \end{tabular}$ &$\kern-\nulldelimiterspace
  \begin{tabular}{@{}c@{}c@{}}
    2.60\\[0.1cm]
    1.07\\[0.1cm]
    0.71\\[0.1cm]
  \end{tabular}$ \\
%% ================================================================================================
%% ================================================================================================
 \kern-\nulldelimiterspace
  \begin{tabular}{@{}c@{}c@{}}
                  \\[0.1cm]
    $^{248}$Cm(1) \\[0.1cm]
                  \\[0.1cm]
  \end{tabular}
 & $\kern-\nulldelimiterspace\left\{
  \begin{tabular}{@{}c@{}c@{}}
    A:\\[0.1cm]
    B:\\[0.1cm]
    C:\\[0.1cm]
  \end{tabular}\right.$
  & $\kern-\nulldelimiterspace
  \begin{tabular}{@{}c@{}c@{}}
    0.0400\\[0.1cm]
    -0.0119	\\[0.1cm]
    $2.99 \times 10^{-3}$\\[0.1cm]
  \end{tabular}$ &$\kern-\nulldelimiterspace
  \begin{tabular}{@{}c@{}c@{}}
    6.675\\[0.1cm]
    7.449\\[0.1cm]
    7.159\\[0.1cm]
  \end{tabular}$ &$\kern-\nulldelimiterspace
  \begin{tabular}{@{}c@{}c@{}}
    $2.682\times10^{-4}$\\[0.1cm]
    $4.747 \times10^{-4}$\\[0.1cm]
    $3.341 \times10^{-4}$\\[0.1cm]
  \end{tabular}$ &$\kern-\nulldelimiterspace
  \begin{tabular}{@{}c@{}c@{}}
                         \\[0.1cm]
    $-1.138\times10^{-7}$\\[0.1cm]
        $8.416 \times10^{-8}$ \\[0.1cm]
  \end{tabular}$ &$\kern-\nulldelimiterspace
  \begin{tabular}{@{}c@{}c@{}}
    \\[0.1cm]
    \\[0.1cm]
  $-9.807 \times10^{-11}$\\[0.1cm]
  \end{tabular}$ &$\kern-\nulldelimiterspace
  \begin{tabular}{@{}c@{}c@{}}
    40.8\\[0.1cm]
    11.4\\[0.1cm]
    1.60\\[0.1cm]
  \end{tabular}$ \\
%% ================================================================================================
%% ================================================================================================
 \kern-\nulldelimiterspace
  \begin{tabular}{@{}c@{}c@{}}
                  \\[0.1cm]
    $^{248}$Cm(2) \\[0.1cm]
                  \\[0.1cm]
  \end{tabular}
 & $\kern-\nulldelimiterspace\left\{
  \begin{tabular}{@{}c@{}c@{}}
    A:\\[0.1cm]
    B:\\[0.1cm]
    C:\\[0.1cm]
  \end{tabular}\right.$
  & $\kern-\nulldelimiterspace
  \begin{tabular}{@{}c@{}c@{}}
    0.0254\\[0.1cm]
    -0.260	\\[0.1cm]
    -0.159\\[0.1cm]
  \end{tabular}$ &$\kern-\nulldelimiterspace
  \begin{tabular}{@{}c@{}c@{}}
    5.326\\[0.1cm]
    7.142\\[0.1cm]
    6.400\\[0.1cm]
  \end{tabular}$ &$\kern-\nulldelimiterspace
  \begin{tabular}{@{}c@{}c@{}}
    $1.275\times10^{-4}$\\[0.1cm]
    $3.616 \times10^{-4}$\\[0.1cm]
    $2.259 \times10^{-4}$\\[0.1cm]
  \end{tabular}$ &$\kern-\nulldelimiterspace
  \begin{tabular}{@{}c@{}c@{}}
                         \\[0.1cm]
    $-9.266 \times10^{-8}$\\[0.1cm]
        $2.569 \times10^{-8}$ \\[0.1cm]
  \end{tabular}$ &$\kern-\nulldelimiterspace
  \begin{tabular}{@{}c@{}c@{}}
    \\[0.1cm]
    \\[0.1cm]
  $-4.601 \times10^{-11}$\\[0.1cm]
  \end{tabular}$ &$\kern-\nulldelimiterspace
  \begin{tabular}{@{}c@{}c@{}}
    4.39\\[0.1cm]
    0.71\\[0.1cm]
    0.60\\[0.1cm]
  \end{tabular}$ \\
%% ================================================================================================
%% ================================================================================================
 \kern-\nulldelimiterspace
  \begin{tabular}{@{}c@{}c@{}}
                  \\[0.1cm]
    $^{248}$Cm(3) \\[0.1cm]
                  \\[0.1cm]
  \end{tabular}
 & $\kern-\nulldelimiterspace\left\{
  \begin{tabular}{@{}c@{}c@{}}
    A:\\[0.1cm]
    B:\\[0.1cm]
    C:\\[0.1cm]
  \end{tabular}\right.$
  & $\kern-\nulldelimiterspace
  \begin{tabular}{@{}c@{}c@{}}
    -0.2180\\[0.1cm]
    -0.2860	\\[0.1cm]
    0.0567\\[0.1cm]
  \end{tabular}$ &$\kern-\nulldelimiterspace
  \begin{tabular}{@{}c@{}c@{}}
    7.244\\[0.1cm]
    7.712\\[0.1cm]
    5.087\\[0.1cm]
  \end{tabular}$ &$\kern-\nulldelimiterspace
  \begin{tabular}{@{}c@{}c@{}}
    $3.762\times10^{-4}$\\[0.1cm]
    $4.605 \times10^{-4}$\\[0.1cm]
    $-4.502 \times10^{-4}$\\[0.1cm]
  \end{tabular}$ &$\kern-\nulldelimiterspace
  \begin{tabular}{@{}c@{}c@{}}
                         \\[0.1cm]
    $-4.893 \times10^{-8}$\\[0.1cm]
        $1.239 \times10^{-6}$ \\[0.1cm]
  \end{tabular}$ &$\kern-\nulldelimiterspace
  \begin{tabular}{@{}c@{}c@{}}
    \\[0.1cm]
    \\[0.1cm]
  $-7.309 \times10^{-10}$\\[0.1cm]
  \end{tabular}$ &$\kern-\nulldelimiterspace
  \begin{tabular}{@{}c@{}c@{}}
    8.14\\[0.1cm]
    8.50\\[0.1cm]
    5.70\\[0.1cm]
  \end{tabular}$ \\
%% ================================================================================================
%% ================================================================================================
 \kern-\nulldelimiterspace
  \begin{tabular}{@{}c@{}c@{}}
                  \\[0.1cm]
    $^{248}$Cm(4) \\[0.1cm]
                  \\[0.1cm]
  \end{tabular}
 & $\kern-\nulldelimiterspace\left\{
  \begin{tabular}{@{}c@{}c@{}}
    A:\\[0.1cm]
    B:\\[0.1cm]
    C:\\[0.1cm]
  \end{tabular}\right.$
  & $\kern-\nulldelimiterspace
  \begin{tabular}{@{}c@{}c@{}}
    -0.2465\\[0.1cm]
    -0.1419	\\[0.1cm]
    -0.0188\\[0.1cm]
  \end{tabular}$ &$\kern-\nulldelimiterspace
  \begin{tabular}{@{}c@{}c@{}}
    6.233\\[0.1cm]
    5.618\\[0.1cm]
    4.787\\[0.1cm]
  \end{tabular}$ &$\kern-\nulldelimiterspace
  \begin{tabular}{@{}c@{}c@{}}
    $1.813\times10^{-4}$\\[0.1cm]
    $4.762 \times10^{-5}$\\[0.1cm]
    $-2.800 \times10^{-4}$\\[0.1cm]
  \end{tabular}$ &$\kern-\nulldelimiterspace
  \begin{tabular}{@{}c@{}c@{}}
                         \\[0.1cm]
    $8.556\times10^{-8}$\\[0.1cm]
        $5.706 \times10^{-7}$ \\[0.1cm]
  \end{tabular}$ &$\kern-\nulldelimiterspace
  \begin{tabular}{@{}c@{}c@{}}
    \\[0.1cm]
    \\[0.1cm]
  $-2.903 \times10^{-10}$\\[0.1cm]
  \end{tabular}$ &$\kern-\nulldelimiterspace
  \begin{tabular}{@{}c@{}c@{}}
    3.27\\[0.1cm]
    1.30\\[0.1cm]
    0.34\\[0.1cm]
  \end{tabular}$ \\
%% ================================================================================================
%% ======================================================================================
\hhline{========}
\end{tabular}
\end{table*}
%%%%%%%%%%%%%%%%%%%%%%%%%%%%%%%%%%%%%%%%%%%%%%%%%%%%%%%%%%%%%%%%%%%%%%%%%%%%%%%%%%%%%%%%%%
% ===================================================================
\begin{figure}
\centering

   \includegraphics[width=8cm]{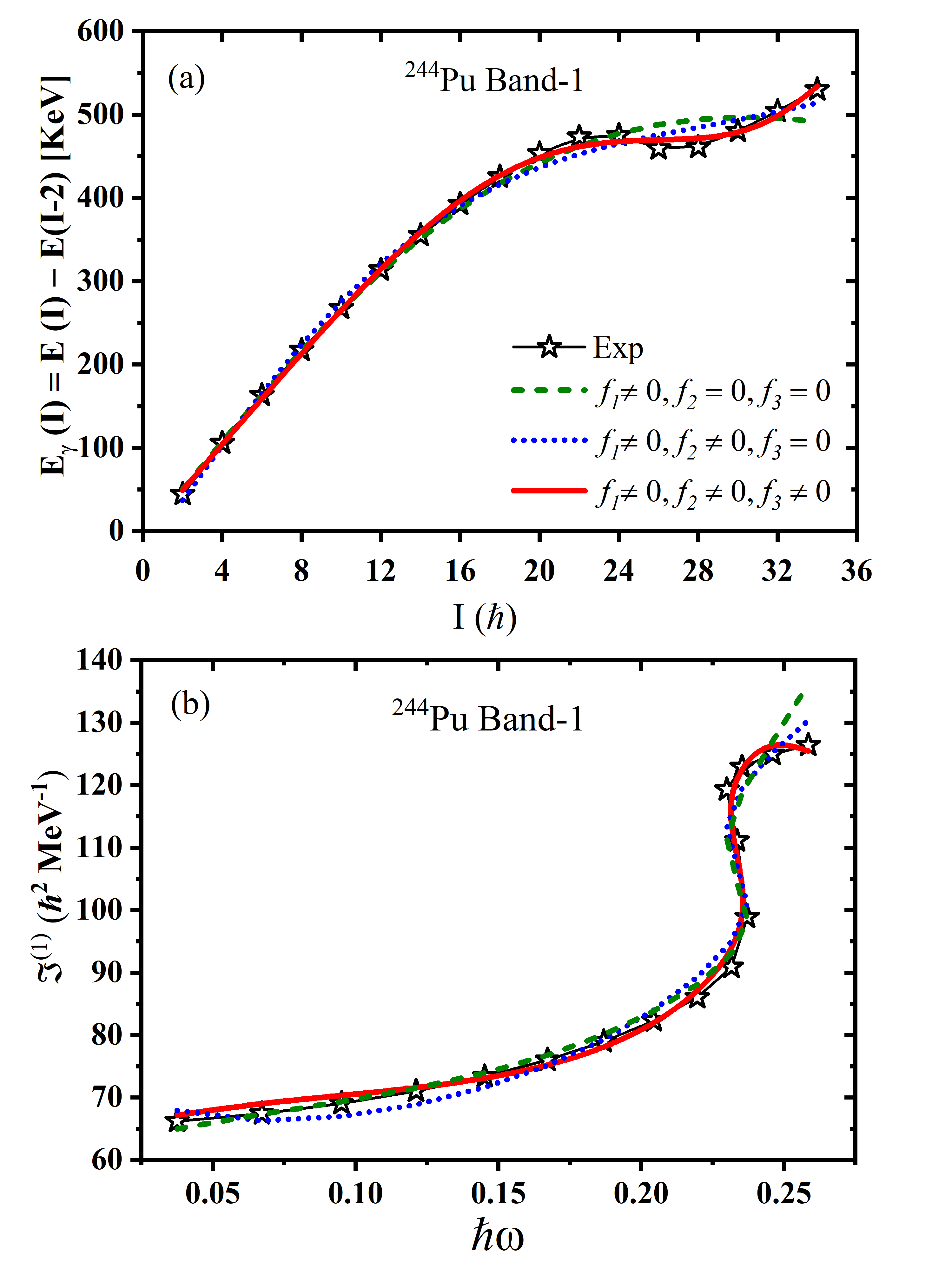}
   \captionsetup{justification=raggedright, singlelinecheck=false}
\caption{(a) The figure provides a comparative analysis between the experimental and calculated intraband gamma-transition energies ($E_\gamma(I)$) plotted against spin, and (b) illustrates the relationship between the kinematic moment of inertia ($\Im^{(1)}$) and the rotational frequency ($\hbar\omega$) for the isotope $^{244}$Pu(1).}
\label{fig1}
\end{figure}
% ===================================================================
% ===================================================================
\begin{figure*}
\centering

   \includegraphics[width=1\textwidth]{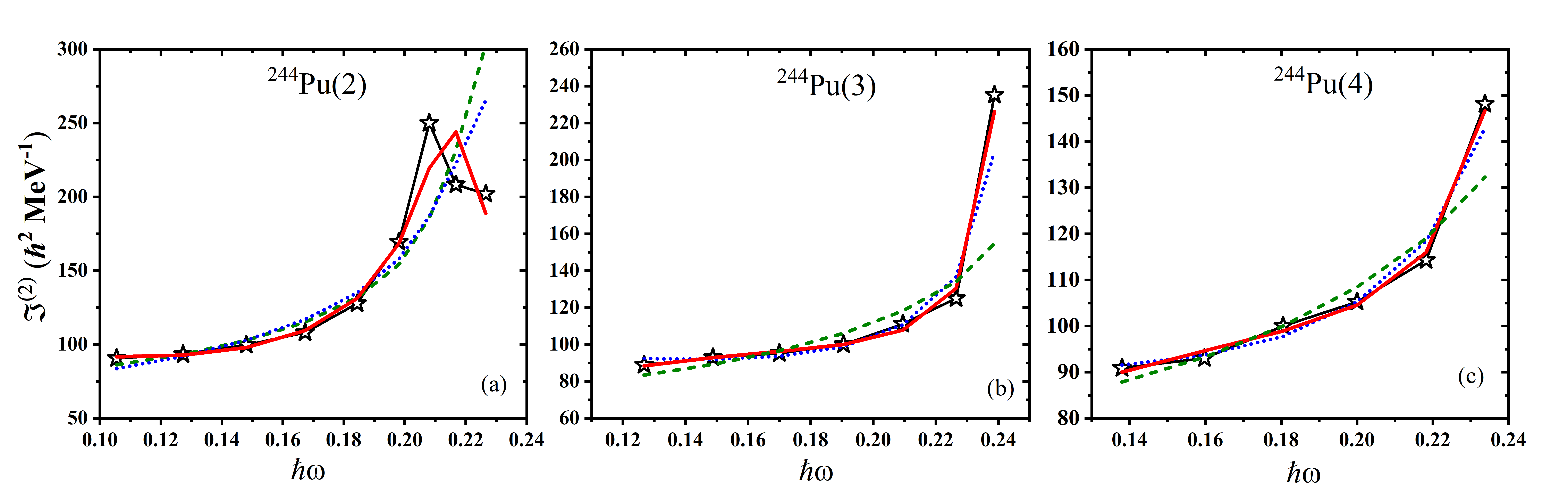}
   \captionsetup{justification=raggedright, singlelinecheck=false}
\caption{The comparison of experimental and calculated dynamic moment of inertia ($\Im^{(2)}$) $vs.$ rotational frequency $\hbar\omega$, for rotational bands in $^{244}$Pu.}
\label{fig2}
\end{figure*}
% ===================================================================
% ===================================================================
\begin{figure*}
\centering

   \includegraphics[width=0.9\textwidth]{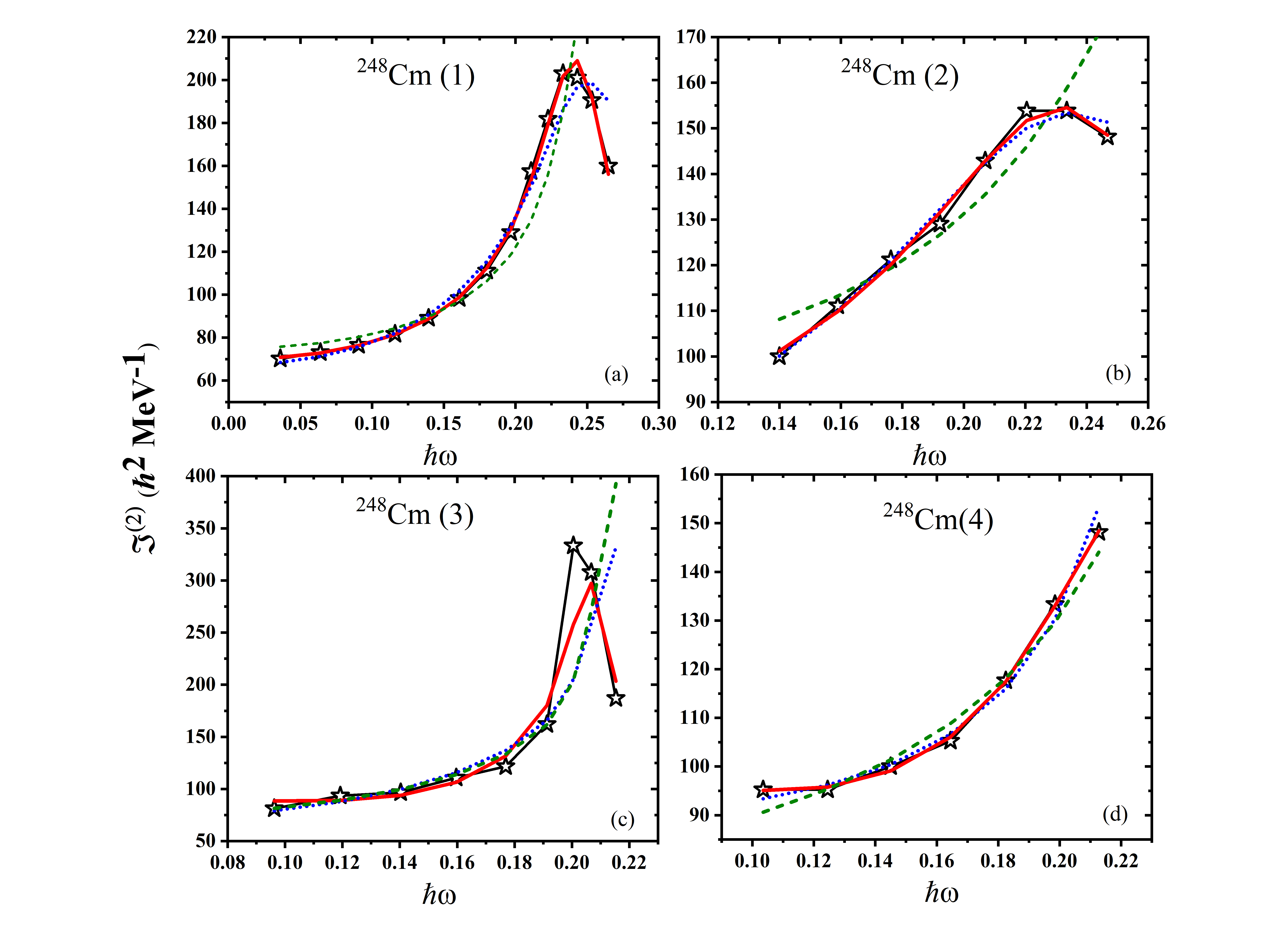}
   \captionsetup{justification=raggedright, singlelinecheck=false}
\caption{The comparison of experimental and calculated dynamic moment of inertia ($\Im^{(2)}$) $vs.$ rotational frequency $\hbar\omega$, for rotational bands in $^{248}$Cm.}
\label{fig3}
\end{figure*}
% ===================================================================
% ===================================================================
\begin{figure*}
\centering

   \includegraphics[width=1\textwidth]{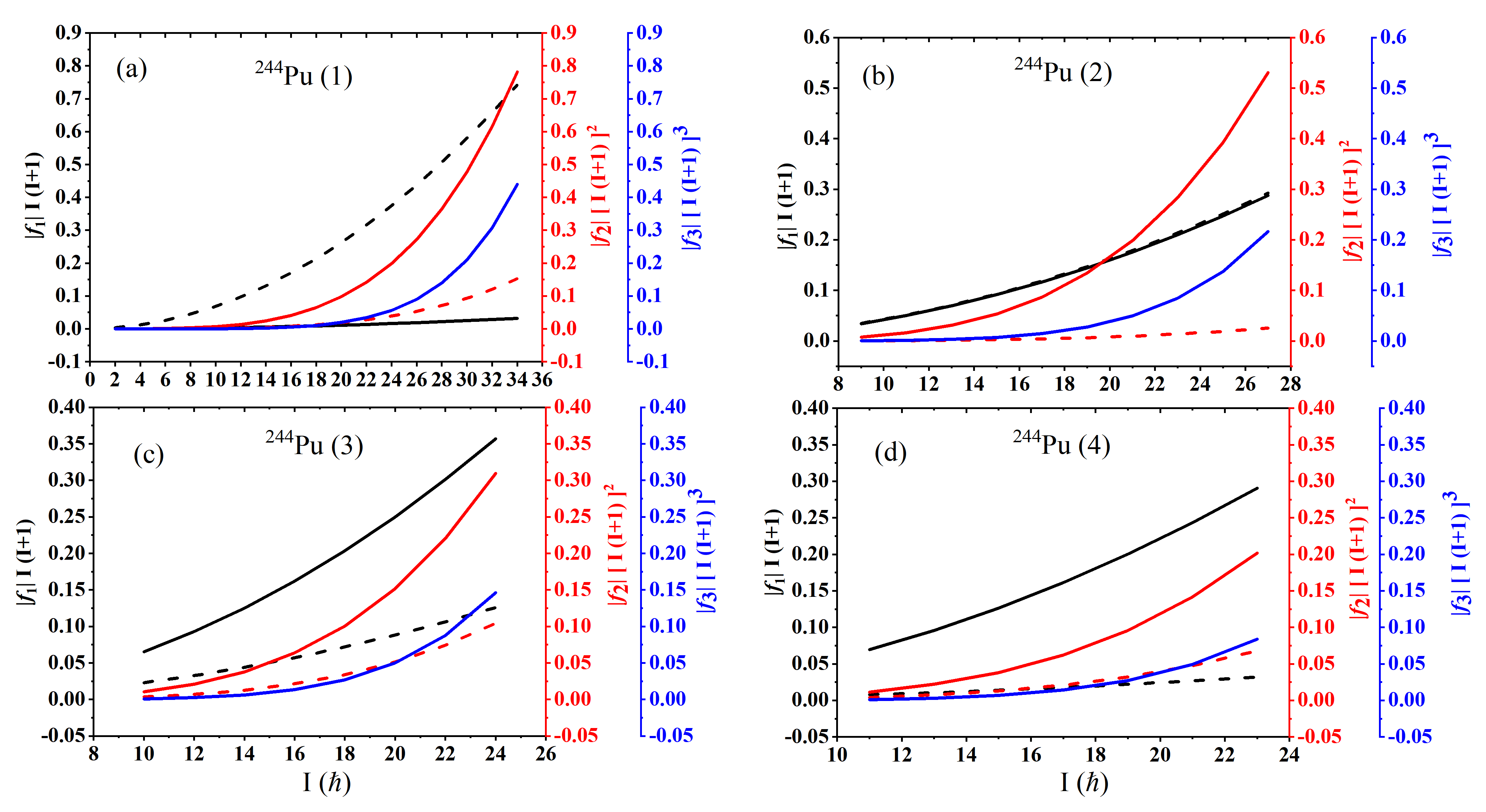}
   \captionsetup{justification=raggedright, singlelinecheck=false}
\caption{This figure presents a comparative analysis of the spin-dependent variations in the parameters $|f_{1}|[I(I+1)]$, $|f_{2}|[I(I+1)]^2$, and $|f_{3}|[I(I+1)]^3$ associated with the rotational bands of $^{244}$Pu, depicted on the black, red, and blue Y-axes respectively.}
\label{fig4}
\end{figure*}
% ===================================================================
%%%%%%%%%%%%%%%%%%%%%%%%%%%%%%%%%%%%%%%%%%%%%%%%%%%%%%%%%%%%%%%%%%%%%%%%%%%
\section{Results and Discussion}
In this paper, we focus on two even-even nuclei, $^{244}$Pu and $^{248}$Cm, as subjects of our study. Both nuclei exhibit rotational bands characterized by notable variations, including back-bending, up-bending, and downturn, in the kinematic/dynamic moment of inertia (MoI) at higher frequency regions \cite{Spreng,HerzbergCox+2011+441+457}. These distinct features render $^{244}$Pu and $^{248}$Cm ideal candidates for investigating the effectiveness of the VMI-inspired IBM within the higher $A\sim250$ mass region.\par
%%%%%%%%%%%%%%%%%%%%%%%%%%%%%%%%%%%%%%%%%%%%
The irrep are determined from the Eq. \ref{eq5}. Illustratively, with the branching rules of the irreps, we get \((\tau_1, \tau_2)\) \(= (16,\, 2),\, (16,\, 0),\, (14,\, 2),\, (14,\, 0), \ldots\)for $^{244}$Pu band-1 with level sequence \(I = 34, 32, 30, 28, \ldots\) \cite{nndc}. The $E_2$ transitions are taken from Ref.\cite{nndc}. After a non-linear least squares fitting method of the experimental intraband $\gamma-$transition, the respective parameters, and calculated intraband $\gamma-$transitions are obtained. The best fit parameters obtained for $^{244}$Pu and $^{248}$Cm are shown in table \ref{tb1}. For every band, three sets of parameters are deduced in this study. In the current analysis, three separate parameter sets are derived for each rotational band studied. Set A is formulated based on the sole influence of the coefficient $f_1$. Set B is expanded to include the effects of two parameters, $f_1$ and $f_2$. In contrast, Set C is the most inclusive, encompassing the combined contributions of $f_1$, $f_2$, and $f_3$. The table also shows the
root-mean-squares (RMS) deviation obtained in all the cases mentioned. The results clearly indicate that, among the three parameter sets examined in this study, Set C consistently yields the lowest RMS deviation between the calculated and experimental $E_2$ values across all the bands considered. This observation emphasizes the improved precision of Set C in accurately representing the details of the $E_2$ transitions within these bands. Figure \ref{fig1}(a) presents a graphical representation of the intraband $\gamma-$transition energies in $^{244}$Pu for band-1, incorporating both experimental observations and theoretically calculated. This particular rotational band is henceforth designated as $^{244}$Pu(1), and similar nomenclature will be followed for subsequent bands. In the same figure, three distinct theoretical curves are depicted, each corresponding to different calculation scenarios. In the first scenario,
labeled as Case-I, the computation utilizes solely the parameter $f_1$, while setting $f_2$ and $f_3$ to zero. Case-II extends this model by incorporating both $f_1$ and $f_2$ parameters in the calculation. Finally, Case-III advances the model further by employing an extrapolation of the Arima coefficient and integrating the $f_3$ parameter, in conjunction with $f_1$ and $f_2$. Figure \ref{fig1}(a) clearly indicates that in Cases I and II, where the parameter $f_3$ is set to zero, there is a notable discrepancy between the calculated and experimental $\gamma-$transition energies for $^{244}$Pu(1), particularly in the region of higher spin ($I\geq20\hbar$). Conversely, the introduction of a non-zero $f_3$ parameter significantly enhances the agreement with the experimental data, leading to an excellent reproduction of the $\gamma-$transitions. Furthermore, a careful examination of the parameters pertaining to $^{244}$Pu(1), as listed in table \ref{tb1}, reveals that the RMS deviation attains its minimum value for parameter set C. Additionally, it is observed that incorporating the $f_2$ parameter into the calculation results in an increase in the RMS
deviation, which escalates from $46.6 \times 10^{-3}$ to $49.5 \times 10^{-3}$. This indicates that the inclusion of the $f_2$ parameter, rather than enhancing the accuracy of the model, actually leads to a slight decrease in its precision in reproducing the experimental data. Figure \ref{fig1}(b) displays the kinematic MoI, $\Im^{(1)}$, for $^{244}$Pu(1). The data presented in this figure clearly demonstrate that the inclusion of a non-zero parameter $f_3$ significantly improves the model's ability to describe the backbending phenomenon observed in $^{244}$Pu(1). In contrast, Cases I and II, which do not account for $f_3$, fail to accurately replicate the experimental curve, particularly in regions of higher frequency. In figure \ref{fig2}(a), the dynamic Moment of Inertia (MoI) $\Im^{(2)}$ for the band $^{244}$Pu(2) is depicted, showcasing a distinctive downturn at higher rotational frequencies. From this figure, it is evident that Cases I and II, which exclude the $f_3$ parameter, inadequately reproduce this downturn, displaying instead a monotonic increase in $\Im^{(2)}$ with rotational frequency. In stark contrast, Case III, with the inclusion of $f_3\neq0$, effectively mirrors the observed behavior, satisfactorily replicating the downturn in the dynamic MoI. Additionally, the parameter dynamics for $^{244}$Pu(2) show similarities to those of $^{244}$Pu(1), as detailed in table \ref{tb1}. Notably, the RMS deviation reaches its lowest value with parameter set C. Furthermore, the introduction of the $f_2$ parameter into the model results in an increase in the RMS deviation compared to set B, suggesting a less optimal fit for the dynamic MoI of $^{244}$Pu(2) when $f_2$ is included.
%The figure \ref{fig2}(b,c) shows the dynamic
%MoI of $^{244}$Pu(3) and $^{244}$Pu(4). These two bands do not show a downturn in dynamic MoI but
%show an upbending at $\hbar\omega\approx 0.22$ MeV. At first, one could assume that the inclusion
%of parameter $f_{1}$ could be sufficient to produce the upbending in the dynamic MoI because it
%incorporates the anti-pairing effect if $f_{1}>0$. However, it can be seen from the figure \ref{fig2}
%that the dynamic MoI calculated from cases I and II is less in magnitude at the highest rotational
%frequency, and only case III reproduces it satisfactorily. \par
Figures \ref{fig2}(b) and \ref{fig2}(c) illustrate the dynamic Moment of Inertia (MoI) for the bands $^{244}$Pu(3) and $^{244}$Pu(4), respectively. Contrary to the previous band, these two do not exhibit a downturn in their dynamic MoI. Instead, they display an upbending phenomenon at a rotational frequency of approximately $\hbar\omega \approx 0.22$ MeV. Initially, one might conjecture that incorporating the parameter $f_{1}$ alone could adequately reproduce this upbending in the dynamic MoI, particularly since $f_{1}$, when positive, accounts for the anti-pairing effect. However, a closer examination of figure \ref{fig2} reveals that in Cases I and II, where the calculation includes only $f_{1}$ or both $f_{1}$ and $f_{2}$ but excludes $f_{3}$, the magnitude of the dynamic MoI at the highest rotational frequencies is underestimated. It is only in Case III, which integrates the $f_{3}$ parameter, that the dynamic MoI for these bands is reproduced with satisfactory accuracy.\par
%%%%%%%%%%%%%%%%%%%%%%%%%%%%%%%%%%%%%%%%%%%%%%%%%%%%%%%%%%%%%%%%%%%%%%%%%%%%%%%%%%%%%
The methodology utilized for $^{244}$Pu has been similarly applied to the $^{248}$Cm nucleus. Within $^{248}$Cm, bands 1, 2, and 3
display a downturn in their dynamic Moment of Inertia (MoI), while band 4 is distinguished by an upbending, as illustrated in
figures \ref{fig3}(a)-(d). These figures indicate that Case III, which incorporates the $f_3$ parameter, accurately reflects the
experimental values for the dynamic MoI in $^{248}$Cm bands 1, 2, and 4. For $^{248}$Cm(3), Case III manages to capture the
overall trend of the dynamic MoI, albeit with less precision. Notably, in $^{248}$Cm(2), the downturn in the dynamic MoI is not
as pronounced as in bands 1 and 3. In this scenario, both Case II and Case III provide satisfactory representations of the dynamic
MoI. However, it is Case III that achieves a closer match to the experimental data. This pattern is also evident in $^{248}$Cm(4).
Here, the experimental data are better represented by Case III as opposed to Case II. This consistency in accurately modeling the
dynamic MoI across various bands of $^{244}$Pu and $^{248}$Cm underscores the importance and effectiveness of including the $f_3$ parameter, especially for capturing detailed phenomena such as backbending, downturns, and up-bendings in the dynamic MoI.\par
%%%%%%%%%%%%%%%%%%%%%%%%%%%%%%%%%%%%%%%%%%%%%%%%%%%%%%%%%%%%%%%%%%%%%%%%%%%%%%%%%%%%%%%
As outlined in Ref. \cite{YOSHIDA}, the incorporation of the $f_1$ parameter in the Hamiltonian is pivotal for modeling pairing and
anti-pairing effects in nuclear systems. Specifically, a positive $f_1$ ($f_1>0$) induces an anti-pairing effect, whereas a negative
$f_1$ ($f_1<0$) facilitates a pairing effect. Extending this concept, it has been recognized that the anti-pairing effect is
intensified when both $f_1$ and $f_2$ are positive ($f_1>0, f_2>0$), and conversely, the pairing effect is strengthened when
both parameters are negative ($f_1<0, f_2<0$). When $f_1$ and $f_2$ assume opposite signs (either $f_1<0, f_2>0$ or $f_1>0, f_2<0$),
both anti-pairing and pairing effects become influential in determining the evolution of the dynamic MoI with rotational frequency. In scenarios where $f_1>0$ and $f_2<0$, there is a shift from an anti-pairing-dominated regime (where angular momentum is the driving factor) to one favoring pairing (characterized by a restraining influence) as angular momentum increases. Conversely, when $f_1<0$ and $f_2>0$, the system transitions from a pairing-dominated regime (restraining) to one favoring anti-pairing (angular momentum driving) with increasing angular momentum. This intricate interplay and the resulting shifts between pairing and anti-pairing effects, dictated by the values of $f_1$ and $f_2$, are extensively discussed in literature \cite{Yuxin_JPG,yuxin_prc_1,yuxin_prc_2,yuxin_prc_3,yuxin_prc_4,yuxin_prc_5,yuxin_prc_6}. The pairing effects plays a vital role before the turnover in the dynamic MoI appears. To further explore the contribution of the parameters $f_1$ and $f_2$, and the importance
of the inclusion of parameter $f_3$ in reproducing the dynamic MoI of $A\sim250$ mass region, we have plotted the variation of
parameters with spins. For $^{244}$Pu(1), the figure \ref{fig4}(a) shows the variation of $|f_1|I(I+1)$, $|f_2|[I(I+1)]^{2}$ and $|f_3|[I(I+1)]^3$ with spin. The analysis clearly shows that \( |f_1| \) is significantly larger than \( |f_2| \), which in turn is considerably greater than \( |f_3| \), indicated by the relationships \( |f_1|\gg|f_2|\gg|f_3| \) (see table \ref{tb1}). This disparity in magnitudes reveals that the contributions of \( f_2 \) and \( f_3 \) become notably significant only at higher spin levels.
%For figure\ref{fig4}, the solid-lines represent the parameter calculated when $f_3$ is included in the calculations
%and the dashed-lines represents the parameters variation when $f_3=0$. The black and red lines (solid and dashed) represents the
%contribution from parameter $f_1$ and $f_2$, respectively and solid blue line represent contribution from $f_3$. For comparing the
%parameters on the same scale, we have plotted the absolute values and ignored the signs. However, it should be noted that the parameters $f_1$, $f_2$
%and $f_3$ have positive and negative values based on different dynamic MoI (see table \ref{tb1}).
% ===================================================================
\begin{figure*}
\centering

   \includegraphics[width=1\textwidth]{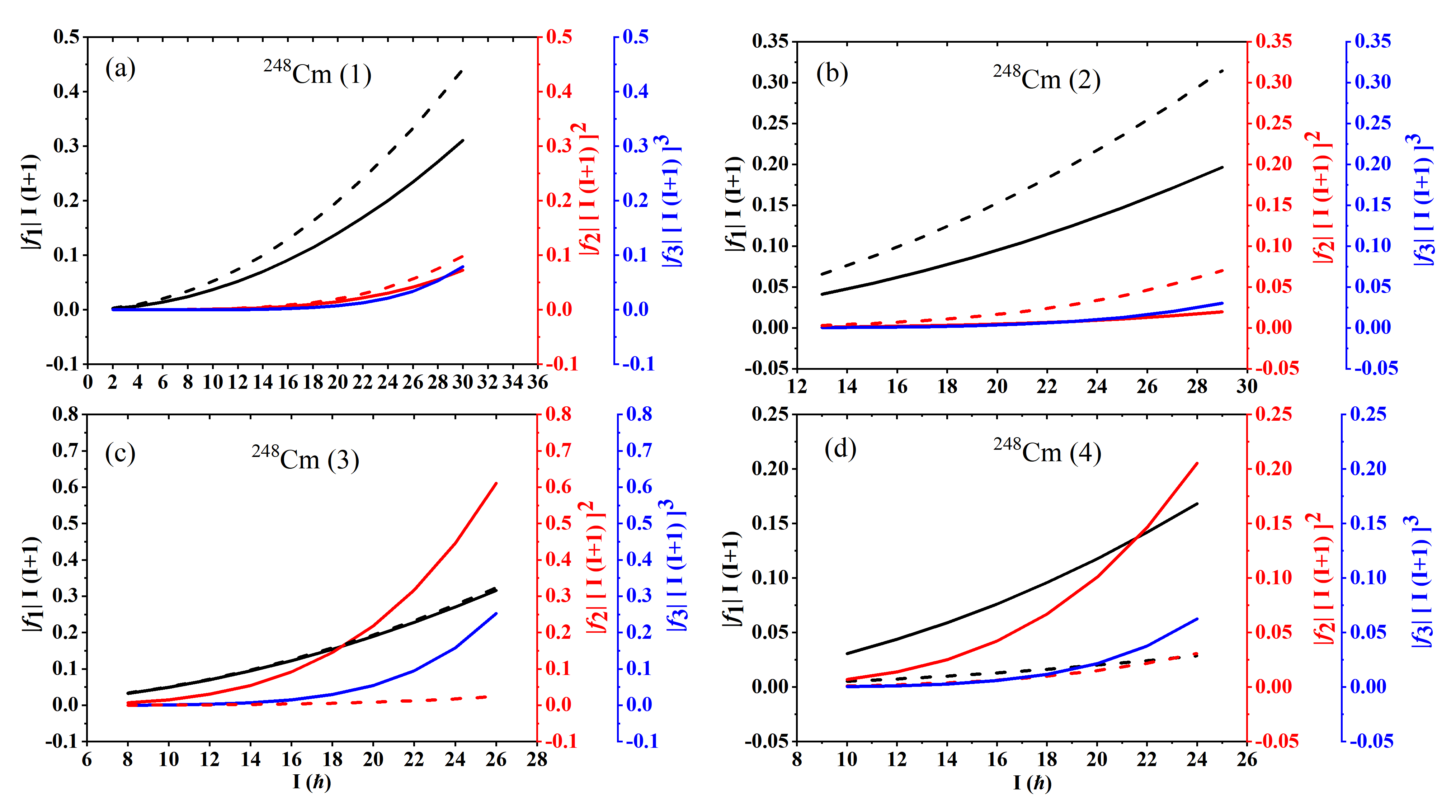}
   \captionsetup{justification=raggedright, singlelinecheck=false}
\caption{This figure presents a comparative analysis of the spin-dependent variations in the parameters $|f_{1}|[I(I+1)]$, $|f_{2}|[I(I+1)]^2$, and $|f_{3}|[I(I+1)]^3$ associated with the rotational bands of  $^{248}$Cm, depicted on the black, red, and blue Y-axes respectively.}
\label{fig5}
\end{figure*}
% ===================================================================
In figure \ref{fig4}, the depiction of parameters is differentiated by the style and color of the lines. The solid lines represent the calculated parameters when the \( f_3 \) factor is included, whereas the dashed lines indicate the variations of these parameters under the condition that \( f_3=0 \). In terms of color coding, the black and red lines, both solid and dashed, correspond to the contributions from the parameters \( f_1 \) and \( f_2 \), respectively. The solid blue line exclusively represents the contribution from \( f_3 \). To facilitate a direct comparison of these parameters on a unified scale, the absolute values of these parameters are plotted, deliberately omitting the signs. This approach allows for a clear visual comparison of the magnitude of each parameter's contribution. However, it is important to note that the actual values of parameters \( f_1 \), \( f_2 \), and \( f_3 \) can vary
between positive and negative, depending on the specific dynamics of the MoI being analyzed, as detailed in table \ref{tb1}. This variation in sign is significant, as it reflects the differing influences these parameters have under various nuclear dynamic conditions. When scrutinizing the parameters, it is imperative to carefully consider the sign each parameter bears, as it is indicative of distinct nuclear effects. Specifically, a parameter with positive sign, denoting the anti-pairing effect, implies a tendency towards reducing the pairing correlations, which is often associated with states of higher excitation or energy within the nucleus. In contrast, a negative sign indicates a pairing favoring effect, suggesting an enhancement or support for pairing correlations. This is typically linked with more stable nuclear states, often found at lower energy levels.
 %For
%$^{244}$Pu(1), parameters $f_1$ and $f_2$ represents the anti-pairing favouring term and $f_3$ is the pairing favouring term. It can
%be seen from the figure\ref{fig4}(a) that $f_1$ term contribution to anti-paring effect is almost negligible throughout the spin range. The
%parameter $f_2$ has negligible contribution up to $\approx10\hbar$ and become higher in magnitude in $I>10\hbar$.
%For pairing favouring term $f_3$, it can be seen that it only become significant after $I>20\hbar$. At maximum spin value $34\hbar$,
%the paring favouring term is almost 1/2 of anti-pairing favouring terms. When compared with the results for $f_3=0$ parameters,
%it is quite clear that parameter $f_1$ (dashed black curve) is dominating over parameter $f_2$ (dashed red curve). With $f_3=0$,
%the parameters $f_1$ and $f_2$ has positive and negative values, respectively (see table \ref{tb1}) which implies the domination of anti-pairing
%favouring effect in $^{244}$Pu(1). However, the comparison of $f_3=0$ and $f_3\neq0$ parameters clear shows the importance of
%increased pairing favouring term in reproducing the experimental data for $^{244}$Pu(1).\par
In the analysis of $^{244}$Pu(1), the roles of parameters $f_1$, $f_2$, and $f_3$ in determining nuclear dynamics, especially in
terms of anti-pairing and pairing effects, are evident from the data presented in figure \ref{fig4}(a). The $f_1$ term, associated
with the anti-pairing effect, shows a negligible contribution across the entire range of spins, suggesting its minimal influence on
the nuclear behavior in $^{244}$Pu(1). Conversely, the $f_2$ parameter, another anti-pairing favoring term, demonstrates a negligible
contribution up to approximately $10\hbar$ and becomes significantly more influential at higher spin values. This indicates a growing
impact of the anti-pairing effect as the spin increases. On the other hand, the $f_3$ term, which favors pairing, only becomes
significant after the spin exceeds $20\hbar$. This implies that the pairing favoring effect is particularly crucial at higher spin states. At the maximum spin value of $34\hbar$, the pairing favoring term's magnitude reaches about half of the anti-pairing favoring terms combined, underscoring the substantial role of pairing effects at extreme spin values. Furthermore, a comparison between scenarios with and without the inclusion of $f_3$ reveals distinct dynamics. When $f_3$ is set to zero, the dominance of the $f_1$ parameter over $f_2$ suggests a prevailing anti-pairing effect. The positive and negative values of $f_1$ and $f_2$, respectively, are detailed in table \ref{tb1}. However, including the $f_3$ parameter significantly alters this balance, emphasizing the increased
importance of the pairing favoring term in accurately reproducing the experimental data for $^{244}$Pu(1). The contribution of each parameter thus emerges as not only dependent on its magnitude but also intricately linked to the spin range.\par
%%%%%%%%%%%%%%%%%%%%%%%%%%%%%%%
For the case of $^{244}$Pu(2), the analysis based on the presented figures, particularly figure \ref{fig4}(b), reveals insightful dynamics about the interplay of anti-pairing and pairing effects. When the $f_3$ parameter is set to zero, the system is predominantly influenced by the anti-pairing favoring effect, with the pairing favoring effect being almost negligible throughout the range of spins. This dominance of the anti-pairing effect is reflected in the calculated dynamic MoI, which shows a monotonous increase with the rotational frequency. However, when $f_3$ is not equal to zero, there is a noticeable shift in the dynamics. Both $f_1$ and $f_3$ parameters contribute towards enhancing the pairing effect. This implies that for an accurate global reproduction of the experimental data, the contributions from pairing favoring terms must be significantly higher. This finding underscores the importance of considering the $f_3$ parameter in the model to capture the nuanced behavior of $^{244}$Pu(2). The behavior observed in $^{244}$Pu(3) and $^{244}$Pu(4) demonstrates similar parameter systematics, as indicated in table \ref{tb1}. In the scenarios where $f_3=0$, the parameters $f_1$ and $f_2$ exhibit negative (pairing favoring) and positive (anti-pairing favoring) values, respectively. In contrast, when $f_3$ is included (i.e., $f_3\neq0$), both $f_1$ and $f_3$ are positive, while $f_2$ remains negative. This configuration suggests a balanced contribution from both anti-pairing and pairing effects in the evolution of the dynamic MoI with rotational frequency, indicating a competition between these two effects. Particularly for $^{244}$Pu(3), the inclusion of the $f_3$ parameter seems to strengthen the anti-pairing effect compared to the results obtained when $f_3=0$.\par
%%%%%%%%%%%%%%%%%%%%%%%%%%%%%
In the nuclear structure of $^{248}$Cm, specifically in bands 1 and 2, a systematic examination of the parameters, as shown in table \ref{tb1}, reveals notable trends. In scenarios where $f_3$ is not considered (\(f_3=0\)), $f_1$ exhibits a positive value, while $f_2$ is negative, indicating an interplay of effects. However, the introduction of a non-zero \(f_3\) (\(f_3\neq0\)) results in both $f_1$ and $f_2$ maintaining their positive values, but with $f_3$ assuming a negative value. This adjustment, as discernible from figures \ref{fig5}(a) and (b), leads to a marked reduction in the anti-pairing favoring term, highlighting the critical role of $f_3$ in the model. The impact of these parameters becomes even more pronounced in $^{248}$Cm(3). In the absence of $f_3$, the anti-pairing effect predominates, leading to a continuous increase in the dynamic MoI, as observed in figure \ref{fig5}(c). However, the integration of a non-zero $f_3$ shifts the balance, bringing a significant contribution from the pairing favoring term. This results in a complex interplay between the pairing and anti-pairing effects, which is pivotal in the evolution of the dynamic MoI. For $^{248}$Cm(4), with $f_3$ set to zero, both $f_1$ and $f_2$ parameters are positive, suggesting an amplified anti-pairing effect. The incorporation of the $f_3$ term, however, not only aligns the theoretical model more closely with experimental data but also introduces a noticeable pairing favoring term.
\section{Summary}
The systematic investigation of the dynamic moment of inertia (MoI) in rotational bands of even-even nuclei $^{244}$Pu and $^{248}$Cm has been conducted using a refined approach of the variable moment of inertia model, inspired by the interacting boson model. This approach incorporates a perturbed $SU_{sdg}(3)$ symmetry integrated with an interaction upholding $SO_{sgd}(5)$ ($SU_{sdg}(5)$) symmetry. A significant advancement is made by extending the Arima coefficients to include three parameters: $f_1$, $f_2$, and $f_3$. This extension allows the intraband $\gamma$-transition energies to be depicted by a five-parameter formula, consisting of two terms. The first term, $B[\tau_1(\tau_1 +3)+\tau_2(\tau_2 +1)]$, retains the $SO_{sdg}(5)$ symmetry, while the second term, $C_{0}/[1+f_{1}I(I+1)+f_{1}I^2(I+1)^2+f_{3}I^3(I+1)^3] I(I+1)$, exhibits $SO(3)$ symmetry, incorporating many-body interactions.\par
%%%%%%%%%%%%%%%%%%%%%%%%%%%%%%%%%%%%%
In these nuclei, the rotational bands demonstrate various changes in MoI, including back-bending, up-bending, and downturns. A closer analysis of the dynamic MoI reveals that the inclusion of only the $f_1$ parameter yields results significantly lower than experimental data. The introduction of the $f_2$ parameter enhances results beyond those achieved with the exclusive use of the $f_1$ model. Nevertheless, it remains insufficient in accurately replicating the experimental variations observed in MoI. This limitation is addressed by introducing the $f_3$ parameter, which effectively replicates the experimental trends. The inclusion of $f_1$ and $f_2$ alone in the model accounts for pairing and anti-pairing effects. However, with the addition of $f_3$, these effects are considerably amplified, with the pairing effects becoming more pronounced in rotational bands where the dynamic MoI experiences a downturn at higher rotational frequencies. Conversely, anti-pairing effects are intensified in bands exhibiting up-bending in the dynamic MoI. Given that the parameters satisfy the relationship $|f_1|\gg |f_2|\gg |f_3|$, it becomes evident that the impact of $|f_3|$ is particularly significant at higher spin values.\par
%%%%%%%%%%%%%%%%%%%%%%%%%%%%%%%%%%%%%%%%
In conclusion, the introduction of the $f_3$ parameter has led to the identification of three distinct phenomena in the rotational bands of $^{244}$Pu and $^{248}$Cm. In the bands $^{244}$Pu(1,2) and $^{248}$Cm(3), an enhancement in the pairing effect is observed. For the bands $^{244}$Pu(3,4), a dominance of the anti-pairing effect becomes evident. Meanwhile, in the bands $^{248}$Cm(1,2,4), there is a noticeable decrease in the anti-pairing effect. This comprehensive analysis underscores the significance of the $f_3$ parameter in capturing the complex interplay of correlations within the nucleus, crucial for understanding the rotational behavior in these heavy nuclei.
% ===================================================================
\begin{acknowledgements}
XTH would like thank the National Natural Science Foundation of China (Grant Nos. U2032138 and 11775112).
\end{acknowledgements}
% ===================================================================

%\nocite{*}
\bibliography{apssamp}

\providecommand{\noopsort}[1]{}\providecommand{\singleletter}[1]{#1}%
\begin{thebibliography}{29}%
\makeatletter
\providecommand \@ifxundefined [1]{%
 \@ifx{#1\undefined}
}%
\providecommand \@ifnum [1]{%
 \ifnum #1\expandafter \@firstoftwo
 \else \expandafter \@secondoftwo
 \fi
}%
\providecommand \@ifx [1]{%
 \ifx #1\expandafter \@firstoftwo
 \else \expandafter \@secondoftwo
 \fi
}%
\providecommand \natexlab [1]{#1}%
\providecommand \enquote  [1]{``#1''}%
\providecommand \bibnamefont  [1]{#1}%
\providecommand \bibfnamefont [1]{#1}%
\providecommand \citenamefont [1]{#1}%
\providecommand \href@noop [0]{\@secondoftwo}%
\providecommand \href [0]{\begingroup \@sanitize@url \@href}%
\providecommand \@href[1]{\@@startlink{#1}\@@href}%
\providecommand \@@href[1]{\endgroup#1\@@endlink}%
\providecommand \@sanitize@url [0]{\catcode `\\12\catcode `\$12\catcode
  `\&12\catcode `\#12\catcode `\^12\catcode `\_12\catcode `\%12\relax}%
\providecommand \@@startlink[1]{}%
\providecommand \@@endlink[0]{}%
\providecommand \url  [0]{\begingroup\@sanitize@url \@url }%
\providecommand \@url [1]{\endgroup\@href {#1}{\urlprefix }}%
\providecommand \urlprefix  [0]{URL }%
\providecommand \Eprint [0]{\href }%
\providecommand \doibase [0]{http://dx.doi.org/}%
\providecommand \selectlanguage [0]{\@gobble}%
\providecommand \bibinfo  [0]{\@secondoftwo}%
\providecommand \bibfield  [0]{\@secondoftwo}%
\providecommand \translation [1]{[#1]}%
\providecommand \BibitemOpen [0]{}%
\providecommand \bibitemStop [0]{}%
\providecommand \bibitemNoStop [0]{.\EOS\space}%
\providecommand \EOS [0]{\spacefactor3000\relax}%
\providecommand \BibitemShut  [1]{\csname bibitem#1\endcsname}%
\let\auto@bib@innerbib\@empty
%</preamble>
\bibitem [{\citenamefont {Stoyer}(2006)}]{Stoyer2006}%
  \BibitemOpen
  \bibfield  {author} {\bibinfo {author} {\bibfnamefont {M.~A.}\ \bibnamefont
  {Stoyer}},\ }\href {\doibase 10.1038/442876a} {\bibfield  {journal} {\bibinfo
   {journal} {Nature}\ }\textbf {\bibinfo {volume} {442}},\ \bibinfo {pages}
  {876} (\bibinfo {year} {2006})}\BibitemShut {NoStop}%
\bibitem [{nnd()}]{nndc}%
  \BibitemOpen
  \href@noop {} {\enquote {\bibinfo {title} {{National Nuclear Data Center
  }},}\ }\bibinfo {howpublished}
  {\url{https://www.nndc.bnl.gov/chart/}}\BibitemShut {NoStop}%
\bibitem [{\citenamefont {Herzberg}\ and\ \citenamefont
  {Cox}(2011)}]{HerzbergCox+2011+441+457}%
  \BibitemOpen
  \bibfield  {author} {\bibinfo {author} {\bibfnamefont {R.-D.}\ \bibnamefont
  {Herzberg}}\ and\ \bibinfo {author} {\bibfnamefont {D.~M.}\ \bibnamefont
  {Cox}},\ }\href {\doibase doi:10.1524/ract.2011.1858} {\bibfield  {journal}
  {\bibinfo  {journal} {Radiochimica Acta}\ }\textbf {\bibinfo {volume} {99}},\
  \bibinfo {pages} {441} (\bibinfo {year} {2011})}\BibitemShut {NoStop}%
\bibitem [{\citenamefont {Leino}\ and\ \citenamefont
  {He\ss{}berger}(2004)}]{Leino}%
  \BibitemOpen
  \bibfield  {author} {\bibinfo {author} {\bibfnamefont {M.}~\bibnamefont
  {Leino}}\ and\ \bibinfo {author} {\bibfnamefont {F.}~\bibnamefont
  {He\ss{}berger}},\ }\href {\doibase 10.1146/annurev.nucl.53.041002.110332}
  {\bibfield  {journal} {\bibinfo  {journal} {Annual Review of Nuclear and
  Particle Science}\ }\textbf {\bibinfo {volume} {54}},\ \bibinfo {pages} {175}
  (\bibinfo {year} {2004})},\ \Eprint
  {http://arxiv.org/abs/https://doi.org/10.1146/annurev.nucl.53.041002.110332}
  {https://doi.org/10.1146/annurev.nucl.53.041002.110332} \BibitemShut
  {NoStop}%
\bibitem [{\citenamefont {Herzberg}(2004)}]{Herzberg_2004}%
  \BibitemOpen
  \bibfield  {author} {\bibinfo {author} {\bibfnamefont {R.-D.}\ \bibnamefont
  {Herzberg}},\ }\href {\doibase 10.1088/0954-3899/30/4/R01} {\bibfield
  {journal} {\bibinfo  {journal} {Journal of Physics G: Nuclear and Particle
  Physics}\ }\textbf {\bibinfo {volume} {30}},\ \bibinfo {pages} {R123}
  (\bibinfo {year} {2004})}\BibitemShut {NoStop}%
\bibitem [{\citenamefont {Greenlees}(2007)}]{GREENLEES2007507}%
  \BibitemOpen
  \bibfield  {author} {\bibinfo {author} {\bibfnamefont {P.}~\bibnamefont
  {Greenlees}},\ }\href {\doibase
  https://doi.org/10.1016/j.nuclphysa.2006.12.078} {\bibfield  {journal}
  {\bibinfo  {journal} {Nuclear Physics A}\ }\textbf {\bibinfo {volume}
  {787}},\ \bibinfo {pages} {507} (\bibinfo {year} {2007})},\ \bibinfo {note}
  {proceedings of the Ninth International Conference on Nucleus-Nucleus
  Collisions}\BibitemShut {NoStop}%
\bibitem [{\citenamefont {Herzberg}\ and\ \citenamefont
  {Greenlees}(2008)}]{HERZBERG2008674}%
  \BibitemOpen
  \bibfield  {author} {\bibinfo {author} {\bibfnamefont {R.-D.}\ \bibnamefont
  {Herzberg}}\ and\ \bibinfo {author} {\bibfnamefont {P.}~\bibnamefont
  {Greenlees}},\ }\href {\doibase https://doi.org/10.1016/j.ppnp.2008.05.003}
  {\bibfield  {journal} {\bibinfo  {journal} {Progress in Particle and Nuclear
  Physics}\ }\textbf {\bibinfo {volume} {61}},\ \bibinfo {pages} {674}
  (\bibinfo {year} {2008})}\BibitemShut {NoStop}%
\bibitem [{\citenamefont {Reiter}\ \emph {et~al.}(1999)\citenamefont {Reiter},
  \citenamefont {Khoo}, \citenamefont {Lister}, \citenamefont {Seweryniak},
  \citenamefont {Ahmad}, \citenamefont {Alcorta}, \citenamefont {Carpenter},
  \citenamefont {Cizewski}, \citenamefont {Davids}, \citenamefont {Gervais},
  \citenamefont {Greene}, \citenamefont {Henning}, \citenamefont {Janssens},
  \citenamefont {Lauritsen}, \citenamefont {Siem}, \citenamefont {Sonzogni},
  \citenamefont {Sullivan}, \citenamefont {Uusitalo}, \citenamefont
  {Wiedenh\"over}, \citenamefont {Amzal}, \citenamefont {Butler}, \citenamefont
  {Chewter}, \citenamefont {Ding}, \citenamefont {Fotiades}, \citenamefont
  {Fox}, \citenamefont {Greenlees}, \citenamefont {Herzberg}, \citenamefont
  {Jones}, \citenamefont {Korten}, \citenamefont {Leino},\ and\ \citenamefont
  {Vetter}}]{Reiter1}%
  \BibitemOpen
  \bibfield  {author} {\bibinfo {author} {\bibfnamefont {P.}~\bibnamefont
  {Reiter}}, \bibinfo {author} {\bibfnamefont {T.~L.}\ \bibnamefont {Khoo}},
  \bibinfo {author} {\bibfnamefont {C.~J.}\ \bibnamefont {Lister}}, \bibinfo
  {author} {\bibfnamefont {D.}~\bibnamefont {Seweryniak}}, \bibinfo {author}
  {\bibfnamefont {I.}~\bibnamefont {Ahmad}}, \bibinfo {author} {\bibfnamefont
  {M.}~\bibnamefont {Alcorta}}, \bibinfo {author} {\bibfnamefont {M.~P.}\
  \bibnamefont {Carpenter}}, \bibinfo {author} {\bibfnamefont {J.~A.}\
  \bibnamefont {Cizewski}}, \bibinfo {author} {\bibfnamefont {C.~N.}\
  \bibnamefont {Davids}}, \bibinfo {author} {\bibfnamefont {G.}~\bibnamefont
  {Gervais}}, \bibinfo {author} {\bibfnamefont {J.~P.}\ \bibnamefont {Greene}},
  \bibinfo {author} {\bibfnamefont {W.~F.}\ \bibnamefont {Henning}}, \bibinfo
  {author} {\bibfnamefont {R.~V.~F.}\ \bibnamefont {Janssens}}, \bibinfo
  {author} {\bibfnamefont {T.}~\bibnamefont {Lauritsen}}, \bibinfo {author}
  {\bibfnamefont {S.}~\bibnamefont {Siem}}, \bibinfo {author} {\bibfnamefont
  {A.~A.}\ \bibnamefont {Sonzogni}}, \bibinfo {author} {\bibfnamefont
  {D.}~\bibnamefont {Sullivan}}, \bibinfo {author} {\bibfnamefont
  {J.}~\bibnamefont {Uusitalo}}, \bibinfo {author} {\bibfnamefont
  {I.}~\bibnamefont {Wiedenh\"over}}, \bibinfo {author} {\bibfnamefont
  {N.}~\bibnamefont {Amzal}}, \bibinfo {author} {\bibfnamefont {P.~A.}\
  \bibnamefont {Butler}}, \bibinfo {author} {\bibfnamefont {A.~J.}\
  \bibnamefont {Chewter}}, \bibinfo {author} {\bibfnamefont {K.~Y.}\
  \bibnamefont {Ding}}, \bibinfo {author} {\bibfnamefont {N.}~\bibnamefont
  {Fotiades}}, \bibinfo {author} {\bibfnamefont {J.~D.}\ \bibnamefont {Fox}},
  \bibinfo {author} {\bibfnamefont {P.~T.}\ \bibnamefont {Greenlees}}, \bibinfo
  {author} {\bibfnamefont {R.-D.}\ \bibnamefont {Herzberg}}, \bibinfo {author}
  {\bibfnamefont {G.~D.}\ \bibnamefont {Jones}}, \bibinfo {author}
  {\bibfnamefont {W.}~\bibnamefont {Korten}}, \bibinfo {author} {\bibfnamefont
  {M.}~\bibnamefont {Leino}}, \ and\ \bibinfo {author} {\bibfnamefont
  {K.}~\bibnamefont {Vetter}},\ }\href {\doibase 10.1103/PhysRevLett.82.509}
  {\bibfield  {journal} {\bibinfo  {journal} {Phys. Rev. Lett.}\ }\textbf
  {\bibinfo {volume} {82}},\ \bibinfo {pages} {509} (\bibinfo {year}
  {1999})}\BibitemShut {NoStop}%
\bibitem [{\citenamefont {Reiter}\ \emph {et~al.}(2000)\citenamefont {Reiter},
  \citenamefont {Khoo}, \citenamefont {Lauritsen}, \citenamefont {Lister},
  \citenamefont {Seweryniak}, \citenamefont {Sonzogni}, \citenamefont {Ahmad},
  \citenamefont {Amzal}, \citenamefont {Bhattacharyya}, \citenamefont {Butler},
  \citenamefont {Carpenter}, \citenamefont {Chewter}, \citenamefont {Cizewski},
  \citenamefont {Davids}, \citenamefont {Ding}, \citenamefont {Fotiades},
  \citenamefont {Greene}, \citenamefont {Greenlees}, \citenamefont {Heinz},
  \citenamefont {Henning}, \citenamefont {Herzberg}, \citenamefont {Janssens},
  \citenamefont {Jones}, \citenamefont {Kondev}, \citenamefont {Korten},
  \citenamefont {Leino}, \citenamefont {Siem}, \citenamefont {Uusitalo},
  \citenamefont {Vetter},\ and\ \citenamefont {Wiedenh\"over}}]{Reiter2}%
  \BibitemOpen
  \bibfield  {author} {\bibinfo {author} {\bibfnamefont {P.}~\bibnamefont
  {Reiter}}, \bibinfo {author} {\bibfnamefont {T.~L.}\ \bibnamefont {Khoo}},
  \bibinfo {author} {\bibfnamefont {T.}~\bibnamefont {Lauritsen}}, \bibinfo
  {author} {\bibfnamefont {C.~J.}\ \bibnamefont {Lister}}, \bibinfo {author}
  {\bibfnamefont {D.}~\bibnamefont {Seweryniak}}, \bibinfo {author}
  {\bibfnamefont {A.~A.}\ \bibnamefont {Sonzogni}}, \bibinfo {author}
  {\bibfnamefont {I.}~\bibnamefont {Ahmad}}, \bibinfo {author} {\bibfnamefont
  {N.}~\bibnamefont {Amzal}}, \bibinfo {author} {\bibfnamefont
  {P.}~\bibnamefont {Bhattacharyya}}, \bibinfo {author} {\bibfnamefont {P.~A.}\
  \bibnamefont {Butler}}, \bibinfo {author} {\bibfnamefont {M.~P.}\
  \bibnamefont {Carpenter}}, \bibinfo {author} {\bibfnamefont {A.~J.}\
  \bibnamefont {Chewter}}, \bibinfo {author} {\bibfnamefont {J.~A.}\
  \bibnamefont {Cizewski}}, \bibinfo {author} {\bibfnamefont {C.~N.}\
  \bibnamefont {Davids}}, \bibinfo {author} {\bibfnamefont {K.~Y.}\
  \bibnamefont {Ding}}, \bibinfo {author} {\bibfnamefont {N.}~\bibnamefont
  {Fotiades}}, \bibinfo {author} {\bibfnamefont {J.~P.}\ \bibnamefont
  {Greene}}, \bibinfo {author} {\bibfnamefont {P.~T.}\ \bibnamefont
  {Greenlees}}, \bibinfo {author} {\bibfnamefont {A.}~\bibnamefont {Heinz}},
  \bibinfo {author} {\bibfnamefont {W.~F.}\ \bibnamefont {Henning}}, \bibinfo
  {author} {\bibfnamefont {R.-D.}\ \bibnamefont {Herzberg}}, \bibinfo {author}
  {\bibfnamefont {R.~V.~F.}\ \bibnamefont {Janssens}}, \bibinfo {author}
  {\bibfnamefont {G.~D.}\ \bibnamefont {Jones}}, \bibinfo {author}
  {\bibfnamefont {F.~G.}\ \bibnamefont {Kondev}}, \bibinfo {author}
  {\bibfnamefont {W.}~\bibnamefont {Korten}}, \bibinfo {author} {\bibfnamefont
  {M.}~\bibnamefont {Leino}}, \bibinfo {author} {\bibfnamefont
  {S.}~\bibnamefont {Siem}}, \bibinfo {author} {\bibfnamefont {J.}~\bibnamefont
  {Uusitalo}}, \bibinfo {author} {\bibfnamefont {K.}~\bibnamefont {Vetter}}, \
  and\ \bibinfo {author} {\bibfnamefont {I.}~\bibnamefont {Wiedenh\"over}},\
  }\href {\doibase 10.1103/PhysRevLett.84.3542} {\bibfield  {journal} {\bibinfo
   {journal} {Phys. Rev. Lett.}\ }\textbf {\bibinfo {volume} {84}},\ \bibinfo
  {pages} {3542} (\bibinfo {year} {2000})}\BibitemShut {NoStop}%
\bibitem [{\citenamefont {Reiter}\ \emph {et~al.}(2005)\citenamefont {Reiter},
  \citenamefont {Khoo}, \citenamefont {Ahmad}, \citenamefont {Afanasjev},
  \citenamefont {Heinz}, \citenamefont {Lauritsen}, \citenamefont {Lister},
  \citenamefont {Seweryniak}, \citenamefont {Bhattacharyya}, \citenamefont
  {Butler}, \citenamefont {Carpenter}, \citenamefont {Chewter}, \citenamefont
  {Cizewski}, \citenamefont {Davids}, \citenamefont {Greene}, \citenamefont
  {Greenlees}, \citenamefont {Helariutta}, \citenamefont {Herzberg},
  \citenamefont {Janssens}, \citenamefont {Jones}, \citenamefont {Julin},
  \citenamefont {Kankaanp\"a\"a}, \citenamefont {Kettunen}, \citenamefont
  {Kondev}, \citenamefont {Kuusiniemi}, \citenamefont {Leino}, \citenamefont
  {Siem}, \citenamefont {Sonzogni}, \citenamefont {Uusitalo},\ and\
  \citenamefont {Wiedenh\"over}}]{Reiter3}%
  \BibitemOpen
  \bibfield  {author} {\bibinfo {author} {\bibfnamefont {P.}~\bibnamefont
  {Reiter}}, \bibinfo {author} {\bibfnamefont {T.~L.}\ \bibnamefont {Khoo}},
  \bibinfo {author} {\bibfnamefont {I.}~\bibnamefont {Ahmad}}, \bibinfo
  {author} {\bibfnamefont {A.~V.}\ \bibnamefont {Afanasjev}}, \bibinfo {author}
  {\bibfnamefont {A.}~\bibnamefont {Heinz}}, \bibinfo {author} {\bibfnamefont
  {T.}~\bibnamefont {Lauritsen}}, \bibinfo {author} {\bibfnamefont {C.~J.}\
  \bibnamefont {Lister}}, \bibinfo {author} {\bibfnamefont {D.}~\bibnamefont
  {Seweryniak}}, \bibinfo {author} {\bibfnamefont {P.}~\bibnamefont
  {Bhattacharyya}}, \bibinfo {author} {\bibfnamefont {P.~A.}\ \bibnamefont
  {Butler}}, \bibinfo {author} {\bibfnamefont {M.~P.}\ \bibnamefont
  {Carpenter}}, \bibinfo {author} {\bibfnamefont {A.~J.}\ \bibnamefont
  {Chewter}}, \bibinfo {author} {\bibfnamefont {J.~A.}\ \bibnamefont
  {Cizewski}}, \bibinfo {author} {\bibfnamefont {C.~N.}\ \bibnamefont
  {Davids}}, \bibinfo {author} {\bibfnamefont {J.~P.}\ \bibnamefont {Greene}},
  \bibinfo {author} {\bibfnamefont {P.~T.}\ \bibnamefont {Greenlees}}, \bibinfo
  {author} {\bibfnamefont {K.}~\bibnamefont {Helariutta}}, \bibinfo {author}
  {\bibfnamefont {R.-D.}\ \bibnamefont {Herzberg}}, \bibinfo {author}
  {\bibfnamefont {R.~V.~F.}\ \bibnamefont {Janssens}}, \bibinfo {author}
  {\bibfnamefont {G.~D.}\ \bibnamefont {Jones}}, \bibinfo {author}
  {\bibfnamefont {R.}~\bibnamefont {Julin}}, \bibinfo {author} {\bibfnamefont
  {H.}~\bibnamefont {Kankaanp\"a\"a}}, \bibinfo {author} {\bibfnamefont
  {H.}~\bibnamefont {Kettunen}}, \bibinfo {author} {\bibfnamefont {F.~G.}\
  \bibnamefont {Kondev}}, \bibinfo {author} {\bibfnamefont {P.}~\bibnamefont
  {Kuusiniemi}}, \bibinfo {author} {\bibfnamefont {M.}~\bibnamefont {Leino}},
  \bibinfo {author} {\bibfnamefont {S.}~\bibnamefont {Siem}}, \bibinfo {author}
  {\bibfnamefont {A.~A.}\ \bibnamefont {Sonzogni}}, \bibinfo {author}
  {\bibfnamefont {J.}~\bibnamefont {Uusitalo}}, \ and\ \bibinfo {author}
  {\bibfnamefont {I.}~\bibnamefont {Wiedenh\"over}},\ }\href {\doibase
  10.1103/PhysRevLett.95.032501} {\bibfield  {journal} {\bibinfo  {journal}
  {Phys. Rev. Lett.}\ }\textbf {\bibinfo {volume} {95}},\ \bibinfo {pages}
  {032501} (\bibinfo {year} {2005})}\BibitemShut {NoStop}%
\bibitem [{\citenamefont {Humphreys}\ \emph {et~al.}(2004)\citenamefont
  {Humphreys}, \citenamefont {Butler}, \citenamefont {Bastin}, \citenamefont
  {Greenlees}, \citenamefont {Hammond}, \citenamefont {Herzberg}, \citenamefont
  {Jenkins}, \citenamefont {Jones}, \citenamefont {Kankaanp\"a\"a},
  \citenamefont {Keenan}, \citenamefont {Kettunen}, \citenamefont {Page},
  \citenamefont {Rahkila}, \citenamefont {Scholey}, \citenamefont {Uusitalo},
  \citenamefont {Amzal}, \citenamefont {Brew}, \citenamefont {Eskola},
  \citenamefont {Gerl}, \citenamefont {Hauschild}, \citenamefont {Helariutta},
  \citenamefont {He\ss{}berger}, \citenamefont {H\"urstel}, \citenamefont
  {Jones}, \citenamefont {Julin}, \citenamefont {Juutinen}, \citenamefont
  {Khoo}, \citenamefont {Korten}, \citenamefont {Kuusiniemi}, \citenamefont
  {Le~Coz}, \citenamefont {Leino}, \citenamefont {Lepp\"anen}, \citenamefont
  {Muikku}, \citenamefont {Nieminen}, \citenamefont {\O{}deg\aa{}rd},
  \citenamefont {Pakarinen}, \citenamefont {Reiter}, \citenamefont {Sletten},
  \citenamefont {Theisen},\ and\ \citenamefont {Wollersheim}}]{Hump}%
  \BibitemOpen
  \bibfield  {author} {\bibinfo {author} {\bibfnamefont {R.~D.}\ \bibnamefont
  {Humphreys}}, \bibinfo {author} {\bibfnamefont {P.~A.}\ \bibnamefont
  {Butler}}, \bibinfo {author} {\bibfnamefont {J.~E.}\ \bibnamefont {Bastin}},
  \bibinfo {author} {\bibfnamefont {P.~T.}\ \bibnamefont {Greenlees}}, \bibinfo
  {author} {\bibfnamefont {N.~J.}\ \bibnamefont {Hammond}}, \bibinfo {author}
  {\bibfnamefont {R.-D.}\ \bibnamefont {Herzberg}}, \bibinfo {author}
  {\bibfnamefont {D.~G.}\ \bibnamefont {Jenkins}}, \bibinfo {author}
  {\bibfnamefont {G.~D.}\ \bibnamefont {Jones}}, \bibinfo {author}
  {\bibfnamefont {H.}~\bibnamefont {Kankaanp\"a\"a}}, \bibinfo {author}
  {\bibfnamefont {A.}~\bibnamefont {Keenan}}, \bibinfo {author} {\bibfnamefont
  {H.}~\bibnamefont {Kettunen}}, \bibinfo {author} {\bibfnamefont
  {T.}~\bibnamefont {Page}}, \bibinfo {author} {\bibfnamefont {P.}~\bibnamefont
  {Rahkila}}, \bibinfo {author} {\bibfnamefont {C.}~\bibnamefont {Scholey}},
  \bibinfo {author} {\bibfnamefont {J.}~\bibnamefont {Uusitalo}}, \bibinfo
  {author} {\bibfnamefont {N.}~\bibnamefont {Amzal}}, \bibinfo {author}
  {\bibfnamefont {P.~M.}\ \bibnamefont {Brew}}, \bibinfo {author}
  {\bibfnamefont {K.}~\bibnamefont {Eskola}}, \bibinfo {author} {\bibfnamefont
  {J.}~\bibnamefont {Gerl}}, \bibinfo {author} {\bibfnamefont {K.}~\bibnamefont
  {Hauschild}}, \bibinfo {author} {\bibfnamefont {K.}~\bibnamefont
  {Helariutta}}, \bibinfo {author} {\bibfnamefont {F.-P.}\ \bibnamefont
  {He\ss{}berger}}, \bibinfo {author} {\bibfnamefont {A.}~\bibnamefont
  {H\"urstel}}, \bibinfo {author} {\bibfnamefont {P.~M.}\ \bibnamefont
  {Jones}}, \bibinfo {author} {\bibfnamefont {R.}~\bibnamefont {Julin}},
  \bibinfo {author} {\bibfnamefont {S.}~\bibnamefont {Juutinen}}, \bibinfo
  {author} {\bibfnamefont {T.-L.}\ \bibnamefont {Khoo}}, \bibinfo {author}
  {\bibfnamefont {W.}~\bibnamefont {Korten}}, \bibinfo {author} {\bibfnamefont
  {P.}~\bibnamefont {Kuusiniemi}}, \bibinfo {author} {\bibfnamefont
  {Y.}~\bibnamefont {Le~Coz}}, \bibinfo {author} {\bibfnamefont
  {M.}~\bibnamefont {Leino}}, \bibinfo {author} {\bibfnamefont {A.-P.}\
  \bibnamefont {Lepp\"anen}}, \bibinfo {author} {\bibfnamefont
  {M.}~\bibnamefont {Muikku}}, \bibinfo {author} {\bibfnamefont
  {P.}~\bibnamefont {Nieminen}}, \bibinfo {author} {\bibfnamefont {S.~W.}\
  \bibnamefont {\O{}deg\aa{}rd}}, \bibinfo {author} {\bibfnamefont
  {J.}~\bibnamefont {Pakarinen}}, \bibinfo {author} {\bibfnamefont
  {P.}~\bibnamefont {Reiter}}, \bibinfo {author} {\bibfnamefont
  {G.}~\bibnamefont {Sletten}}, \bibinfo {author} {\bibfnamefont
  {C.}~\bibnamefont {Theisen}}, \ and\ \bibinfo {author} {\bibfnamefont
  {H.-J.}\ \bibnamefont {Wollersheim}},\ }\href {\doibase
  10.1103/PhysRevC.69.064324} {\bibfield  {journal} {\bibinfo  {journal} {Phys.
  Rev. C}\ }\textbf {\bibinfo {volume} {69}},\ \bibinfo {pages} {064324}
  (\bibinfo {year} {2004})}\BibitemShut {NoStop}%
\bibitem [{\citenamefont {Butler}\ \emph {et~al.}(2002)\citenamefont {Butler},
  \citenamefont {Humphreys}, \citenamefont {Greenlees}, \citenamefont
  {Herzberg}, \citenamefont {Jenkins}, \citenamefont {Jones}, \citenamefont
  {Kankaanp\"a\"a}, \citenamefont {Kettunen}, \citenamefont {Rahkila},
  \citenamefont {Scholey}, \citenamefont {Uusitalo}, \citenamefont {Amzal},
  \citenamefont {Bastin}, \citenamefont {Brew}, \citenamefont {Eskola},
  \citenamefont {Gerl}, \citenamefont {Hammond}, \citenamefont {Hauschild},
  \citenamefont {Helariutta}, \citenamefont {He\ss{}berger}, \citenamefont
  {H\"urstel}, \citenamefont {Jones}, \citenamefont {Julin}, \citenamefont
  {Juutinen}, \citenamefont {Keenan}, \citenamefont {Khoo}, \citenamefont
  {Korten}, \citenamefont {Kuusiniemi}, \citenamefont {Le~Coz}, \citenamefont
  {Leino}, \citenamefont {Lepp\"anen}, \citenamefont {Muikku}, \citenamefont
  {Nieminen}, \citenamefont {\O{}deg\aa{}rd}, \citenamefont {Page},
  \citenamefont {Pakarinen}, \citenamefont {Reiter}, \citenamefont {Sletten},
  \citenamefont {Theisen},\ and\ \citenamefont {Wollersheim}}]{Butler}%
  \BibitemOpen
  \bibfield  {author} {\bibinfo {author} {\bibfnamefont {P.~A.}\ \bibnamefont
  {Butler}}, \bibinfo {author} {\bibfnamefont {R.~D.}\ \bibnamefont
  {Humphreys}}, \bibinfo {author} {\bibfnamefont {P.~T.}\ \bibnamefont
  {Greenlees}}, \bibinfo {author} {\bibfnamefont {R.-D.}\ \bibnamefont
  {Herzberg}}, \bibinfo {author} {\bibfnamefont {D.~G.}\ \bibnamefont
  {Jenkins}}, \bibinfo {author} {\bibfnamefont {G.~D.}\ \bibnamefont {Jones}},
  \bibinfo {author} {\bibfnamefont {H.}~\bibnamefont {Kankaanp\"a\"a}},
  \bibinfo {author} {\bibfnamefont {H.}~\bibnamefont {Kettunen}}, \bibinfo
  {author} {\bibfnamefont {P.}~\bibnamefont {Rahkila}}, \bibinfo {author}
  {\bibfnamefont {C.}~\bibnamefont {Scholey}}, \bibinfo {author} {\bibfnamefont
  {J.}~\bibnamefont {Uusitalo}}, \bibinfo {author} {\bibfnamefont
  {N.}~\bibnamefont {Amzal}}, \bibinfo {author} {\bibfnamefont {J.~E.}\
  \bibnamefont {Bastin}}, \bibinfo {author} {\bibfnamefont {P.~M.~T.}\
  \bibnamefont {Brew}}, \bibinfo {author} {\bibfnamefont {K.}~\bibnamefont
  {Eskola}}, \bibinfo {author} {\bibfnamefont {J.}~\bibnamefont {Gerl}},
  \bibinfo {author} {\bibfnamefont {N.~J.}\ \bibnamefont {Hammond}}, \bibinfo
  {author} {\bibfnamefont {K.}~\bibnamefont {Hauschild}}, \bibinfo {author}
  {\bibfnamefont {K.}~\bibnamefont {Helariutta}}, \bibinfo {author}
  {\bibfnamefont {F.-P.}\ \bibnamefont {He\ss{}berger}}, \bibinfo {author}
  {\bibfnamefont {A.}~\bibnamefont {H\"urstel}}, \bibinfo {author}
  {\bibfnamefont {P.~M.}\ \bibnamefont {Jones}}, \bibinfo {author}
  {\bibfnamefont {R.}~\bibnamefont {Julin}}, \bibinfo {author} {\bibfnamefont
  {S.}~\bibnamefont {Juutinen}}, \bibinfo {author} {\bibfnamefont
  {A.}~\bibnamefont {Keenan}}, \bibinfo {author} {\bibfnamefont {T.-L.}\
  \bibnamefont {Khoo}}, \bibinfo {author} {\bibfnamefont {W.}~\bibnamefont
  {Korten}}, \bibinfo {author} {\bibfnamefont {P.}~\bibnamefont {Kuusiniemi}},
  \bibinfo {author} {\bibfnamefont {Y.}~\bibnamefont {Le~Coz}}, \bibinfo
  {author} {\bibfnamefont {M.}~\bibnamefont {Leino}}, \bibinfo {author}
  {\bibfnamefont {A.-P.}\ \bibnamefont {Lepp\"anen}}, \bibinfo {author}
  {\bibfnamefont {M.}~\bibnamefont {Muikku}}, \bibinfo {author} {\bibfnamefont
  {P.}~\bibnamefont {Nieminen}}, \bibinfo {author} {\bibfnamefont {S.~W.}\
  \bibnamefont {\O{}deg\aa{}rd}}, \bibinfo {author} {\bibfnamefont
  {T.}~\bibnamefont {Page}}, \bibinfo {author} {\bibfnamefont {J.}~\bibnamefont
  {Pakarinen}}, \bibinfo {author} {\bibfnamefont {P.}~\bibnamefont {Reiter}},
  \bibinfo {author} {\bibfnamefont {G.}~\bibnamefont {Sletten}}, \bibinfo
  {author} {\bibfnamefont {C.}~\bibnamefont {Theisen}}, \ and\ \bibinfo
  {author} {\bibfnamefont {H.-J.}\ \bibnamefont {Wollersheim}},\ }\href
  {\doibase 10.1103/PhysRevLett.89.202501} {\bibfield  {journal} {\bibinfo
  {journal} {Phys. Rev. Lett.}\ }\textbf {\bibinfo {volume} {89}},\ \bibinfo
  {pages} {202501} (\bibinfo {year} {2002})}\BibitemShut {NoStop}%
\bibitem [{\citenamefont {Herzberg}\ \emph {et~al.}(2001)\citenamefont
  {Herzberg}, \citenamefont {Amzal}, \citenamefont {Becker}, \citenamefont
  {Butler}, \citenamefont {Chewter}, \citenamefont {Cocks}, \citenamefont
  {Dorvaux}, \citenamefont {Eskola}, \citenamefont {Gerl}, \citenamefont
  {Greenlees}, \citenamefont {Hammond}, \citenamefont {Hauschild},
  \citenamefont {Helariutta}, \citenamefont {He\ss{}berger}, \citenamefont
  {Houry}, \citenamefont {Jones}, \citenamefont {Jones}, \citenamefont {Julin},
  \citenamefont {Juutinen}, \citenamefont {Kankaanp\"a\"a}, \citenamefont
  {Kettunen}, \citenamefont {Khoo}, \citenamefont {Korten}, \citenamefont
  {Kuusiniemi}, \citenamefont {Coz}, \citenamefont {Leino}, \citenamefont
  {Lister}, \citenamefont {Lucas}, \citenamefont {Muikku}, \citenamefont
  {Nieminen}, \citenamefont {Page}, \citenamefont {Rahkila}, \citenamefont
  {Reiter}, \citenamefont {Schlegel}, \citenamefont {Scholey}, \citenamefont
  {Stezowski}, \citenamefont {Theisen}, \citenamefont {Trzaska}, \citenamefont
  {Uusitalo},\ and\ \citenamefont {Wollersheim}}]{Herzberg}%
  \BibitemOpen
  \bibfield  {author} {\bibinfo {author} {\bibfnamefont {R.-D.}\ \bibnamefont
  {Herzberg}}, \bibinfo {author} {\bibfnamefont {N.}~\bibnamefont {Amzal}},
  \bibinfo {author} {\bibfnamefont {F.}~\bibnamefont {Becker}}, \bibinfo
  {author} {\bibfnamefont {P.~A.}\ \bibnamefont {Butler}}, \bibinfo {author}
  {\bibfnamefont {A.~J.~C.}\ \bibnamefont {Chewter}}, \bibinfo {author}
  {\bibfnamefont {J.~F.~C.}\ \bibnamefont {Cocks}}, \bibinfo {author}
  {\bibfnamefont {O.}~\bibnamefont {Dorvaux}}, \bibinfo {author} {\bibfnamefont
  {K.}~\bibnamefont {Eskola}}, \bibinfo {author} {\bibfnamefont
  {J.}~\bibnamefont {Gerl}}, \bibinfo {author} {\bibfnamefont {P.~T.}\
  \bibnamefont {Greenlees}}, \bibinfo {author} {\bibfnamefont {N.~J.}\
  \bibnamefont {Hammond}}, \bibinfo {author} {\bibfnamefont {K.}~\bibnamefont
  {Hauschild}}, \bibinfo {author} {\bibfnamefont {K.}~\bibnamefont
  {Helariutta}}, \bibinfo {author} {\bibfnamefont {F.}~\bibnamefont
  {He\ss{}berger}}, \bibinfo {author} {\bibfnamefont {M.}~\bibnamefont
  {Houry}}, \bibinfo {author} {\bibfnamefont {G.~D.}\ \bibnamefont {Jones}},
  \bibinfo {author} {\bibfnamefont {P.~M.}\ \bibnamefont {Jones}}, \bibinfo
  {author} {\bibfnamefont {R.}~\bibnamefont {Julin}}, \bibinfo {author}
  {\bibfnamefont {S.}~\bibnamefont {Juutinen}}, \bibinfo {author}
  {\bibfnamefont {H.}~\bibnamefont {Kankaanp\"a\"a}}, \bibinfo {author}
  {\bibfnamefont {H.}~\bibnamefont {Kettunen}}, \bibinfo {author}
  {\bibfnamefont {T.~L.}\ \bibnamefont {Khoo}}, \bibinfo {author}
  {\bibfnamefont {W.}~\bibnamefont {Korten}}, \bibinfo {author} {\bibfnamefont
  {P.}~\bibnamefont {Kuusiniemi}}, \bibinfo {author} {\bibfnamefont {Y.~L.}\
  \bibnamefont {Coz}}, \bibinfo {author} {\bibfnamefont {M.}~\bibnamefont
  {Leino}}, \bibinfo {author} {\bibfnamefont {C.~J.}\ \bibnamefont {Lister}},
  \bibinfo {author} {\bibfnamefont {R.}~\bibnamefont {Lucas}}, \bibinfo
  {author} {\bibfnamefont {M.}~\bibnamefont {Muikku}}, \bibinfo {author}
  {\bibfnamefont {P.}~\bibnamefont {Nieminen}}, \bibinfo {author}
  {\bibfnamefont {R.~D.}\ \bibnamefont {Page}}, \bibinfo {author}
  {\bibfnamefont {P.}~\bibnamefont {Rahkila}}, \bibinfo {author} {\bibfnamefont
  {P.}~\bibnamefont {Reiter}}, \bibinfo {author} {\bibfnamefont
  {C.}~\bibnamefont {Schlegel}}, \bibinfo {author} {\bibfnamefont
  {C.}~\bibnamefont {Scholey}}, \bibinfo {author} {\bibfnamefont
  {O.}~\bibnamefont {Stezowski}}, \bibinfo {author} {\bibfnamefont
  {C.}~\bibnamefont {Theisen}}, \bibinfo {author} {\bibfnamefont {W.~H.}\
  \bibnamefont {Trzaska}}, \bibinfo {author} {\bibfnamefont {J.}~\bibnamefont
  {Uusitalo}}, \ and\ \bibinfo {author} {\bibfnamefont {H.~J.}\ \bibnamefont
  {Wollersheim}},\ }\href {\doibase 10.1103/PhysRevC.65.014303} {\bibfield
  {journal} {\bibinfo  {journal} {Phys. Rev. C}\ }\textbf {\bibinfo {volume}
  {65}},\ \bibinfo {pages} {014303} (\bibinfo {year} {2001})}\BibitemShut
  {NoStop}%
\bibitem [{\citenamefont {Bastin}\ \emph {et~al.}(2006)\citenamefont {Bastin},
  \citenamefont {Herzberg}, \citenamefont {Butler}, \citenamefont {Jones},
  \citenamefont {Page}, \citenamefont {Jenkins}, \citenamefont {Amzal},
  \citenamefont {Brew}, \citenamefont {Hammond}, \citenamefont {Humphreys},
  \citenamefont {Ikin}, \citenamefont {Page}, \citenamefont {Greenlees},
  \citenamefont {Jones}, \citenamefont {Julin}, \citenamefont {Juutinen},
  \citenamefont {Kankaanp\"a\"a}, \citenamefont {Keenan}, \citenamefont
  {Kettunen}, \citenamefont {Kuusiniemi}, \citenamefont {Leino}, \citenamefont
  {Lepp\"anen}, \citenamefont {Muikku}, \citenamefont {Nieminen}, \citenamefont
  {Rahkila}, \citenamefont {Scholey}, \citenamefont {Uusitalo}, \citenamefont
  {Bouchez}, \citenamefont {Chatillon}, \citenamefont {H\"urstel},
  \citenamefont {Korten}, \citenamefont {Coz}, \citenamefont {Theisen},
  \citenamefont {Ackermann}, \citenamefont {Gerl}, \citenamefont {Helariutta},
  \citenamefont {Hessberger}, \citenamefont {Schlegel}, \citenamefont
  {Wollerscheim}, \citenamefont {Lach}, \citenamefont {Maj}, \citenamefont
  {Meczynski}, \citenamefont {Styczen}, \citenamefont {Khoo}, \citenamefont
  {Lister}, \citenamefont {Afanasjev}, \citenamefont {Maier}, \citenamefont
  {Reiter}, \citenamefont {Bednarczyc}, \citenamefont {Eskola},\ and\
  \citenamefont {Hauschild}}]{Bastin}%
  \BibitemOpen
  \bibfield  {author} {\bibinfo {author} {\bibfnamefont {J.~E.}\ \bibnamefont
  {Bastin}}, \bibinfo {author} {\bibfnamefont {R.-D.}\ \bibnamefont
  {Herzberg}}, \bibinfo {author} {\bibfnamefont {P.~A.}\ \bibnamefont
  {Butler}}, \bibinfo {author} {\bibfnamefont {G.~D.}\ \bibnamefont {Jones}},
  \bibinfo {author} {\bibfnamefont {R.~D.}\ \bibnamefont {Page}}, \bibinfo
  {author} {\bibfnamefont {D.~G.}\ \bibnamefont {Jenkins}}, \bibinfo {author}
  {\bibfnamefont {N.}~\bibnamefont {Amzal}}, \bibinfo {author} {\bibfnamefont
  {P.~M.~T.}\ \bibnamefont {Brew}}, \bibinfo {author} {\bibfnamefont {N.~J.}\
  \bibnamefont {Hammond}}, \bibinfo {author} {\bibfnamefont {R.~D.}\
  \bibnamefont {Humphreys}}, \bibinfo {author} {\bibfnamefont {P.~J.~C.}\
  \bibnamefont {Ikin}}, \bibinfo {author} {\bibfnamefont {T.}~\bibnamefont
  {Page}}, \bibinfo {author} {\bibfnamefont {P.~T.}\ \bibnamefont {Greenlees}},
  \bibinfo {author} {\bibfnamefont {P.~M.}\ \bibnamefont {Jones}}, \bibinfo
  {author} {\bibfnamefont {R.}~\bibnamefont {Julin}}, \bibinfo {author}
  {\bibfnamefont {S.}~\bibnamefont {Juutinen}}, \bibinfo {author}
  {\bibfnamefont {H.}~\bibnamefont {Kankaanp\"a\"a}}, \bibinfo {author}
  {\bibfnamefont {A.}~\bibnamefont {Keenan}}, \bibinfo {author} {\bibfnamefont
  {H.}~\bibnamefont {Kettunen}}, \bibinfo {author} {\bibfnamefont
  {P.}~\bibnamefont {Kuusiniemi}}, \bibinfo {author} {\bibfnamefont
  {M.}~\bibnamefont {Leino}}, \bibinfo {author} {\bibfnamefont {A.~P.}\
  \bibnamefont {Lepp\"anen}}, \bibinfo {author} {\bibfnamefont
  {M.}~\bibnamefont {Muikku}}, \bibinfo {author} {\bibfnamefont
  {P.}~\bibnamefont {Nieminen}}, \bibinfo {author} {\bibfnamefont
  {P.}~\bibnamefont {Rahkila}}, \bibinfo {author} {\bibfnamefont
  {C.}~\bibnamefont {Scholey}}, \bibinfo {author} {\bibfnamefont
  {J.}~\bibnamefont {Uusitalo}}, \bibinfo {author} {\bibfnamefont
  {E.}~\bibnamefont {Bouchez}}, \bibinfo {author} {\bibfnamefont
  {A.}~\bibnamefont {Chatillon}}, \bibinfo {author} {\bibfnamefont
  {A.}~\bibnamefont {H\"urstel}}, \bibinfo {author} {\bibfnamefont
  {W.}~\bibnamefont {Korten}}, \bibinfo {author} {\bibfnamefont {Y.~L.}\
  \bibnamefont {Coz}}, \bibinfo {author} {\bibfnamefont {C.}~\bibnamefont
  {Theisen}}, \bibinfo {author} {\bibfnamefont {D.}~\bibnamefont {Ackermann}},
  \bibinfo {author} {\bibfnamefont {J.}~\bibnamefont {Gerl}}, \bibinfo {author}
  {\bibfnamefont {K.}~\bibnamefont {Helariutta}}, \bibinfo {author}
  {\bibfnamefont {F.~P.}\ \bibnamefont {Hessberger}}, \bibinfo {author}
  {\bibfnamefont {C.}~\bibnamefont {Schlegel}}, \bibinfo {author}
  {\bibfnamefont {H.~J.}\ \bibnamefont {Wollerscheim}}, \bibinfo {author}
  {\bibfnamefont {M.}~\bibnamefont {Lach}}, \bibinfo {author} {\bibfnamefont
  {A.}~\bibnamefont {Maj}}, \bibinfo {author} {\bibfnamefont {W.}~\bibnamefont
  {Meczynski}}, \bibinfo {author} {\bibfnamefont {J.}~\bibnamefont {Styczen}},
  \bibinfo {author} {\bibfnamefont {T.~L.}\ \bibnamefont {Khoo}}, \bibinfo
  {author} {\bibfnamefont {C.~J.}\ \bibnamefont {Lister}}, \bibinfo {author}
  {\bibfnamefont {A.~V.}\ \bibnamefont {Afanasjev}}, \bibinfo {author}
  {\bibfnamefont {H.~J.}\ \bibnamefont {Maier}}, \bibinfo {author}
  {\bibfnamefont {P.}~\bibnamefont {Reiter}}, \bibinfo {author} {\bibfnamefont
  {P.}~\bibnamefont {Bednarczyc}}, \bibinfo {author} {\bibfnamefont
  {K.}~\bibnamefont {Eskola}}, \ and\ \bibinfo {author} {\bibfnamefont
  {K.}~\bibnamefont {Hauschild}},\ }\href {\doibase 10.1103/PhysRevC.73.024308}
  {\bibfield  {journal} {\bibinfo  {journal} {Phys. Rev. C}\ }\textbf {\bibinfo
  {volume} {73}},\ \bibinfo {pages} {024308} (\bibinfo {year}
  {2006})}\BibitemShut {NoStop}%
\bibitem [{\citenamefont {Wiedenh\"over}\ \emph {et~al.}(1999)\citenamefont
  {Wiedenh\"over}, \citenamefont {Janssens}, \citenamefont {Hackman},
  \citenamefont {Ahmad}, \citenamefont {Greene}, \citenamefont {Amro},
  \citenamefont {Bhattacharyya}, \citenamefont {Carpenter}, \citenamefont
  {Chowdhury}, \citenamefont {Cizewski}, \citenamefont {Cline}, \citenamefont
  {Khoo}, \citenamefont {Lauritsen}, \citenamefont {Lister}, \citenamefont
  {Macchiavelli}, \citenamefont {Nisius}, \citenamefont {Reiter}, \citenamefont
  {Seabury}, \citenamefont {Seweryniak}, \citenamefont {Siem}, \citenamefont
  {Sonzogni}, \citenamefont {Uusitalo},\ and\ \citenamefont
  {Wu}}]{PhysRevLett.83.2143}%
  \BibitemOpen
  \bibfield  {author} {\bibinfo {author} {\bibfnamefont {I.}~\bibnamefont
  {Wiedenh\"over}}, \bibinfo {author} {\bibfnamefont {R.~V.~F.}\ \bibnamefont
  {Janssens}}, \bibinfo {author} {\bibfnamefont {G.}~\bibnamefont {Hackman}},
  \bibinfo {author} {\bibfnamefont {I.}~\bibnamefont {Ahmad}}, \bibinfo
  {author} {\bibfnamefont {J.~P.}\ \bibnamefont {Greene}}, \bibinfo {author}
  {\bibfnamefont {H.}~\bibnamefont {Amro}}, \bibinfo {author} {\bibfnamefont
  {P.~K.}\ \bibnamefont {Bhattacharyya}}, \bibinfo {author} {\bibfnamefont
  {M.~P.}\ \bibnamefont {Carpenter}}, \bibinfo {author} {\bibfnamefont
  {P.}~\bibnamefont {Chowdhury}}, \bibinfo {author} {\bibfnamefont
  {J.}~\bibnamefont {Cizewski}}, \bibinfo {author} {\bibfnamefont
  {D.}~\bibnamefont {Cline}}, \bibinfo {author} {\bibfnamefont {T.~L.}\
  \bibnamefont {Khoo}}, \bibinfo {author} {\bibfnamefont {T.}~\bibnamefont
  {Lauritsen}}, \bibinfo {author} {\bibfnamefont {C.~J.}\ \bibnamefont
  {Lister}}, \bibinfo {author} {\bibfnamefont {A.~O.}\ \bibnamefont
  {Macchiavelli}}, \bibinfo {author} {\bibfnamefont {D.~T.}\ \bibnamefont
  {Nisius}}, \bibinfo {author} {\bibfnamefont {P.}~\bibnamefont {Reiter}},
  \bibinfo {author} {\bibfnamefont {E.~H.}\ \bibnamefont {Seabury}}, \bibinfo
  {author} {\bibfnamefont {D.}~\bibnamefont {Seweryniak}}, \bibinfo {author}
  {\bibfnamefont {S.}~\bibnamefont {Siem}}, \bibinfo {author} {\bibfnamefont
  {A.}~\bibnamefont {Sonzogni}}, \bibinfo {author} {\bibfnamefont
  {J.}~\bibnamefont {Uusitalo}}, \ and\ \bibinfo {author} {\bibfnamefont
  {C.~Y.}\ \bibnamefont {Wu}},\ }\href {\doibase 10.1103/PhysRevLett.83.2143}
  {\bibfield  {journal} {\bibinfo  {journal} {Phys. Rev. Lett.}\ }\textbf
  {\bibinfo {volume} {83}},\ \bibinfo {pages} {2143} (\bibinfo {year}
  {1999})}\BibitemShut {NoStop}%
\bibitem [{\citenamefont {Wang}\ \emph {et~al.}(2009)\citenamefont {Wang},
  \citenamefont {Janssens}, \citenamefont {Carpenter}, \citenamefont {Zhu},
  \citenamefont {Wiedenh\"over}, \citenamefont {Garg}, \citenamefont
  {Frauendorf}, \citenamefont {Nakatsukasa}, \citenamefont {Ahmad},
  \citenamefont {Bernstein}, \citenamefont {Diffenderfer}, \citenamefont
  {Freeman}, \citenamefont {Greene}, \citenamefont {Khoo}, \citenamefont
  {Kondev}, \citenamefont {Larabee}, \citenamefont {Lauritsen}, \citenamefont
  {Lister}, \citenamefont {Meredith}, \citenamefont {Seweryniak}, \citenamefont
  {Teal},\ and\ \citenamefont {Wilson}}]{PhysRevLett.102.122501}%
  \BibitemOpen
  \bibfield  {author} {\bibinfo {author} {\bibfnamefont {X.}~\bibnamefont
  {Wang}}, \bibinfo {author} {\bibfnamefont {R.~V.~F.}\ \bibnamefont
  {Janssens}}, \bibinfo {author} {\bibfnamefont {M.~P.}\ \bibnamefont
  {Carpenter}}, \bibinfo {author} {\bibfnamefont {S.}~\bibnamefont {Zhu}},
  \bibinfo {author} {\bibfnamefont {I.}~\bibnamefont {Wiedenh\"over}}, \bibinfo
  {author} {\bibfnamefont {U.}~\bibnamefont {Garg}}, \bibinfo {author}
  {\bibfnamefont {S.}~\bibnamefont {Frauendorf}}, \bibinfo {author}
  {\bibfnamefont {T.}~\bibnamefont {Nakatsukasa}}, \bibinfo {author}
  {\bibfnamefont {I.}~\bibnamefont {Ahmad}}, \bibinfo {author} {\bibfnamefont
  {A.}~\bibnamefont {Bernstein}}, \bibinfo {author} {\bibfnamefont
  {E.}~\bibnamefont {Diffenderfer}}, \bibinfo {author} {\bibfnamefont {S.~J.}\
  \bibnamefont {Freeman}}, \bibinfo {author} {\bibfnamefont {J.~P.}\
  \bibnamefont {Greene}}, \bibinfo {author} {\bibfnamefont {T.~L.}\
  \bibnamefont {Khoo}}, \bibinfo {author} {\bibfnamefont {F.~G.}\ \bibnamefont
  {Kondev}}, \bibinfo {author} {\bibfnamefont {A.}~\bibnamefont {Larabee}},
  \bibinfo {author} {\bibfnamefont {T.}~\bibnamefont {Lauritsen}}, \bibinfo
  {author} {\bibfnamefont {C.~J.}\ \bibnamefont {Lister}}, \bibinfo {author}
  {\bibfnamefont {B.}~\bibnamefont {Meredith}}, \bibinfo {author}
  {\bibfnamefont {D.}~\bibnamefont {Seweryniak}}, \bibinfo {author}
  {\bibfnamefont {C.}~\bibnamefont {Teal}}, \ and\ \bibinfo {author}
  {\bibfnamefont {P.}~\bibnamefont {Wilson}},\ }\href {\doibase
  10.1103/PhysRevLett.102.122501} {\bibfield  {journal} {\bibinfo  {journal}
  {Phys. Rev. Lett.}\ }\textbf {\bibinfo {volume} {102}},\ \bibinfo {pages}
  {122501} (\bibinfo {year} {2009})}\BibitemShut {NoStop}%
\bibitem [{\citenamefont {Piercey}\ \emph {et~al.}(1981)\citenamefont
  {Piercey}, \citenamefont {Hamilton}, \citenamefont {Ramayya}, \citenamefont
  {Emling}, \citenamefont {Fuchs}, \citenamefont {Grosse}, \citenamefont
  {Schwalm}, \citenamefont {Wollersheim}, \citenamefont {Trautmann},
  \citenamefont {Faessler},\ and\ \citenamefont
  {Ploszajczak}}]{PhysRevLett.46.415}%
  \BibitemOpen
  \bibfield  {author} {\bibinfo {author} {\bibfnamefont {R.~B.}\ \bibnamefont
  {Piercey}}, \bibinfo {author} {\bibfnamefont {J.~H.}\ \bibnamefont
  {Hamilton}}, \bibinfo {author} {\bibfnamefont {A.~V.}\ \bibnamefont
  {Ramayya}}, \bibinfo {author} {\bibfnamefont {H.}~\bibnamefont {Emling}},
  \bibinfo {author} {\bibfnamefont {P.}~\bibnamefont {Fuchs}}, \bibinfo
  {author} {\bibfnamefont {E.}~\bibnamefont {Grosse}}, \bibinfo {author}
  {\bibfnamefont {D.}~\bibnamefont {Schwalm}}, \bibinfo {author} {\bibfnamefont
  {H.~J.}\ \bibnamefont {Wollersheim}}, \bibinfo {author} {\bibfnamefont
  {N.}~\bibnamefont {Trautmann}}, \bibinfo {author} {\bibfnamefont
  {A.}~\bibnamefont {Faessler}}, \ and\ \bibinfo {author} {\bibfnamefont
  {M.}~\bibnamefont {Ploszajczak}},\ }\href {\doibase
  10.1103/PhysRevLett.46.415} {\bibfield  {journal} {\bibinfo  {journal} {Phys.
  Rev. Lett.}\ }\textbf {\bibinfo {volume} {46}},\ \bibinfo {pages} {415}
  (\bibinfo {year} {1981})}\BibitemShut {NoStop}%
\bibitem [{\citenamefont {Piercey}\ \emph {et~al.}(1993)\citenamefont
  {Piercey}, \citenamefont {Michel}, \citenamefont {Grosse}, \citenamefont
  {Emling}, \citenamefont {Schwalm}, \citenamefont {Wollersheim}, \citenamefont
  {Hamilton}, \citenamefont {Ramayya},\ and\ \citenamefont
  {Trautmann}}]{RBPiercey1993}%
  \BibitemOpen
  \bibfield  {author} {\bibinfo {author} {\bibfnamefont {R.~B.}\ \bibnamefont
  {Piercey}}, \bibinfo {author} {\bibfnamefont {C.}~\bibnamefont {Michel}},
  \bibinfo {author} {\bibfnamefont {E.}~\bibnamefont {Grosse}}, \bibinfo
  {author} {\bibfnamefont {H.}~\bibnamefont {Emling}}, \bibinfo {author}
  {\bibfnamefont {D.}~\bibnamefont {Schwalm}}, \bibinfo {author} {\bibfnamefont
  {H.~J.}\ \bibnamefont {Wollersheim}}, \bibinfo {author} {\bibfnamefont
  {J.~H.}\ \bibnamefont {Hamilton}}, \bibinfo {author} {\bibfnamefont {A.~V.}\
  \bibnamefont {Ramayya}}, \ and\ \bibinfo {author} {\bibfnamefont
  {N.}~\bibnamefont {Trautmann}},\ }\href {\doibase 10.1088/0954-3899/19/6/006}
  {\bibfield  {journal} {\bibinfo  {journal} {Journal of Physics G: Nuclear and
  Particle Physics}\ }\textbf {\bibinfo {volume} {19}},\ \bibinfo {pages} {849}
  (\bibinfo {year} {1993})}\BibitemShut {NoStop}%
\bibitem [{\citenamefont {Liu}\ \emph {et~al.}(1998)\citenamefont {Liu},
  \citenamefont {Song}, \citenamefont {zhou Sun}, \citenamefont {jun Wang},\
  and\ \citenamefont {guang Zhao}}]{Yuxin_JPG}%
  \BibitemOpen
  \bibfield  {author} {\bibinfo {author} {\bibfnamefont {Y.}~\bibnamefont
  {Liu}}, \bibinfo {author} {\bibfnamefont {J.}~\bibnamefont {Song}}, \bibinfo
  {author} {\bibfnamefont {H.}~\bibnamefont {zhou Sun}}, \bibinfo {author}
  {\bibfnamefont {J.}~\bibnamefont {jun Wang}}, \ and\ \bibinfo {author}
  {\bibfnamefont {E.}~\bibnamefont {guang Zhao}},\ }\href
  {http://stacks.iop.org/0954-3899/24/i=1/a=015} {\bibfield  {journal}
  {\bibinfo  {journal} {Journal of Physics G: Nuclear and Particle Physics}\
  }\textbf {\bibinfo {volume} {24}},\ \bibinfo {pages} {117} (\bibinfo {year}
  {1998})}\BibitemShut {NoStop}%
\bibitem [{\citenamefont {Liu}\ \emph {et~al.}(1997)\citenamefont {Liu},
  \citenamefont {Song}, \citenamefont {Sun},\ and\ \citenamefont
  {Zhao}}]{yuxin_prc_1}%
  \BibitemOpen
  \bibfield  {author} {\bibinfo {author} {\bibfnamefont {Y.-X.}\ \bibnamefont
  {Liu}}, \bibinfo {author} {\bibfnamefont {J.-g.}\ \bibnamefont {Song}},
  \bibinfo {author} {\bibfnamefont {H.-z.}\ \bibnamefont {Sun}}, \ and\
  \bibinfo {author} {\bibfnamefont {E.-g.}\ \bibnamefont {Zhao}},\ }\href
  {\doibase 10.1103/PhysRevC.56.1370} {\bibfield  {journal} {\bibinfo
  {journal} {Phys. Rev. C}\ }\textbf {\bibinfo {volume} {56}},\ \bibinfo
  {pages} {1370} (\bibinfo {year} {1997})}\BibitemShut {NoStop}%
\bibitem [{\citenamefont {Liu}(1998{\natexlab{a}})}]{yuxin_prc_2}%
  \BibitemOpen
  \bibfield  {author} {\bibinfo {author} {\bibfnamefont {Y.-X.}\ \bibnamefont
  {Liu}},\ }\href {\doibase 10.1103/PhysRevC.58.237} {\bibfield  {journal}
  {\bibinfo  {journal} {Phys. Rev. C}\ }\textbf {\bibinfo {volume} {58}},\
  \bibinfo {pages} {237} (\bibinfo {year} {1998}{\natexlab{a}})}\BibitemShut
  {NoStop}%
\bibitem [{\citenamefont {Liu}(1998{\natexlab{b}})}]{yuxin_prc_3}%
  \BibitemOpen
  \bibfield  {author} {\bibinfo {author} {\bibfnamefont {Y.-X.}\ \bibnamefont
  {Liu}},\ }\href {\doibase 10.1103/PhysRevC.58.900} {\bibfield  {journal}
  {\bibinfo  {journal} {Phys. Rev. C}\ }\textbf {\bibinfo {volume} {58}},\
  \bibinfo {pages} {900} (\bibinfo {year} {1998}{\natexlab{b}})}\BibitemShut
  {NoStop}%
\bibitem [{\citenamefont {Liu}\ \emph {et~al.}(1999)\citenamefont {Liu},
  \citenamefont {Sun},\ and\ \citenamefont {Zhao}}]{yuxin_prc_4}%
  \BibitemOpen
  \bibfield  {author} {\bibinfo {author} {\bibfnamefont {Y.-X.}\ \bibnamefont
  {Liu}}, \bibinfo {author} {\bibfnamefont {D.}~\bibnamefont {Sun}}, \ and\
  \bibinfo {author} {\bibfnamefont {E.-g.}\ \bibnamefont {Zhao}},\ }\href
  {\doibase 10.1103/PhysRevC.59.2511} {\bibfield  {journal} {\bibinfo
  {journal} {Phys. Rev. C}\ }\textbf {\bibinfo {volume} {59}},\ \bibinfo
  {pages} {2511} (\bibinfo {year} {1999})}\BibitemShut {NoStop}%
\bibitem [{\citenamefont {Liu}\ and\ \citenamefont {Gao}(2001)}]{yuxin_prc_5}%
  \BibitemOpen
  \bibfield  {author} {\bibinfo {author} {\bibfnamefont {Y.-X.}\ \bibnamefont
  {Liu}}\ and\ \bibinfo {author} {\bibfnamefont {D.-f.}\ \bibnamefont {Gao}},\
  }\href {\doibase 10.1103/PhysRevC.63.044317} {\bibfield  {journal} {\bibinfo
  {journal} {Phys. Rev. C}\ }\textbf {\bibinfo {volume} {63}},\ \bibinfo
  {pages} {044317} (\bibinfo {year} {2001})}\BibitemShut {NoStop}%
\bibitem [{\citenamefont {Liu}\ \emph {et~al.}(2001)\citenamefont {Liu},
  \citenamefont {Wang},\ and\ \citenamefont {Han}}]{yuxin_prc_6}%
  \BibitemOpen
  \bibfield  {author} {\bibinfo {author} {\bibfnamefont {Y.-X.}\ \bibnamefont
  {Liu}}, \bibinfo {author} {\bibfnamefont {J.-j.}\ \bibnamefont {Wang}}, \
  and\ \bibinfo {author} {\bibfnamefont {Q.-z.}\ \bibnamefont {Han}},\ }\href
  {\doibase 10.1103/PhysRevC.64.064320} {\bibfield  {journal} {\bibinfo
  {journal} {Phys. Rev. C}\ }\textbf {\bibinfo {volume} {64}},\ \bibinfo
  {pages} {064320} (\bibinfo {year} {2001})}\BibitemShut {NoStop}%
\bibitem [{\citenamefont {Yoshida}\ \emph {et~al.}(1991)\citenamefont
  {Yoshida}, \citenamefont {Sagawa}, \citenamefont {Otsuka},\ and\
  \citenamefont {Arima}}]{YOSHIDA}%
  \BibitemOpen
  \bibfield  {author} {\bibinfo {author} {\bibfnamefont {N.}~\bibnamefont
  {Yoshida}}, \bibinfo {author} {\bibfnamefont {H.}~\bibnamefont {Sagawa}},
  \bibinfo {author} {\bibfnamefont {T.}~\bibnamefont {Otsuka}}, \ and\ \bibinfo
  {author} {\bibfnamefont {A.}~\bibnamefont {Arima}},\ }\href {\doibase
  https://doi.org/10.1016/0370-2693(91)90662-A} {\bibfield  {journal} {\bibinfo
   {journal} {Physics Letters B}\ }\textbf {\bibinfo {volume} {256}},\ \bibinfo
  {pages} {129} (\bibinfo {year} {1991})}\BibitemShut {NoStop}%
\bibitem [{\citenamefont {Da-Li}\ and\ \citenamefont
  {Bin-Gang}(2010)}]{ZhangDaLi}%
  \BibitemOpen
  \bibfield  {author} {\bibinfo {author} {\bibfnamefont {Z.}~\bibnamefont
  {Da-Li}}\ and\ \bibinfo {author} {\bibfnamefont {D.}~\bibnamefont
  {Bin-Gang}},\ }\href {\doibase 10.1088/0256-307X/27/6/062101} {\bibfield
  {journal} {\bibinfo  {journal} {Chinese Physics Letters}\ }\textbf {\bibinfo
  {volume} {27}},\ \bibinfo {pages} {062101} (\bibinfo {year}
  {2010})}\BibitemShut {NoStop}%
\bibitem [{\citenamefont {Da-Li}\ \emph {et~al.}(2010)\citenamefont {Da-Li},
  \citenamefont {Jin-Bo},\ and\ \citenamefont {Hong-Ping}}]{ZhangDaLi2}%
  \BibitemOpen
  \bibfield  {author} {\bibinfo {author} {\bibfnamefont {Z.}~\bibnamefont
  {Da-Li}}, \bibinfo {author} {\bibfnamefont {L.}~\bibnamefont {Jin-Bo}}, \
  and\ \bibinfo {author} {\bibfnamefont {C.}~\bibnamefont {Hong-Ping}},\ }\href
  {\doibase 10.1088/0253-6102/53/1/26} {\bibfield  {journal} {\bibinfo
  {journal} {Communications in Theoretical Physics}\ }\textbf {\bibinfo
  {volume} {53}},\ \bibinfo {pages} {121} (\bibinfo {year} {2010})}\BibitemShut
  {NoStop}%
\bibitem [{\citenamefont {Spreng}\ \emph {et~al.}(1983)\citenamefont {Spreng},
  \citenamefont {Azgui}, \citenamefont {Emling}, \citenamefont {Grosse},
  \citenamefont {Kulessa}, \citenamefont {Michel}, \citenamefont {Schwalm},
  \citenamefont {Simon}, \citenamefont {Wollersheim}, \citenamefont {Mutterer},
  \citenamefont {Theobald}, \citenamefont {Moore}, \citenamefont {Trautmann},
  \citenamefont {Egido},\ and\ \citenamefont {Ring}}]{Spreng}%
  \BibitemOpen
  \bibfield  {author} {\bibinfo {author} {\bibfnamefont {W.}~\bibnamefont
  {Spreng}}, \bibinfo {author} {\bibfnamefont {F.}~\bibnamefont {Azgui}},
  \bibinfo {author} {\bibfnamefont {H.}~\bibnamefont {Emling}}, \bibinfo
  {author} {\bibfnamefont {E.}~\bibnamefont {Grosse}}, \bibinfo {author}
  {\bibfnamefont {R.}~\bibnamefont {Kulessa}}, \bibinfo {author} {\bibfnamefont
  {C.}~\bibnamefont {Michel}}, \bibinfo {author} {\bibfnamefont
  {D.}~\bibnamefont {Schwalm}}, \bibinfo {author} {\bibfnamefont {R.~S.}\
  \bibnamefont {Simon}}, \bibinfo {author} {\bibfnamefont {H.~J.}\ \bibnamefont
  {Wollersheim}}, \bibinfo {author} {\bibfnamefont {M.}~\bibnamefont
  {Mutterer}}, \bibinfo {author} {\bibfnamefont {J.~P.}\ \bibnamefont
  {Theobald}}, \bibinfo {author} {\bibfnamefont {M.~S.}\ \bibnamefont {Moore}},
  \bibinfo {author} {\bibfnamefont {N.}~\bibnamefont {Trautmann}}, \bibinfo
  {author} {\bibfnamefont {J.~L.}\ \bibnamefont {Egido}}, \ and\ \bibinfo
  {author} {\bibfnamefont {P.}~\bibnamefont {Ring}},\ }\href {\doibase
  10.1103/PhysRevLett.51.1522} {\bibfield  {journal} {\bibinfo  {journal}
  {Phys. Rev. Lett.}\ }\textbf {\bibinfo {volume} {51}},\ \bibinfo {pages}
  {1522} (\bibinfo {year} {1983})}\BibitemShut {NoStop}%
\end{thebibliography}%
\end{document}